\documentclass[%
 reprint,
superscriptaddress,
 amsmath,amssymb,
 aps,
 prl,
]{revtex4-2}

\usepackage{graphicx}
\usepackage{dcolumn}
\usepackage{bm}
\usepackage{hyperref}
\usepackage[mathlines]{lineno}
\usepackage{comment}

\usepackage{dcolumn}
\usepackage{multirow}
\usepackage{booktabs}
\usepackage{xcolor}

\newcommand{\Ameas}{A_{PV}^{\rm meas}}
\newcommand{\APV}{A_{\rm PV}}

\begin{document}

\preprint{APS/123-QED}

\title{Accurate Determination of the Neutron Skin Thickness of $^{208}$Pb through Parity-Violation in Electron Scattering}

\collaboration{The PREX Collaboration}

\author{D.~Adhikari}\affiliation{Idaho State University, Pocatello, Idaho 83209, USA}
\author{H.~Albataineh}\affiliation{Texas  A  \&  M  University - Kingsville,  Kingsville,  Texas  78363,  USA}
\author{D.~Androic}\affiliation{University of Zagreb, Faculty of Science, Zagreb, HR 10002, Croatia}
\author{K.~Aniol}\affiliation{California State University, Los Angeles, Los Angeles, California  90032, USA}
\author{D.S.~Armstrong}\affiliation{William \& Mary, Williamsburg, Virginia 23185, USA}
\author{T.~Averett}\affiliation{William \& Mary, Williamsburg, Virginia 23185, USA}
\author{\mbox{C. Ayerbe Gayoso}}\affiliation{William \& Mary, Williamsburg, Virginia 23185, USA}
\author{S.~Barcus}\affiliation{Thomas Jefferson National Accelerator Facility, Newport News, Virginia 23606, USA} 
\author{V.~Bellini}\affiliation{Istituto  Nazionale  di  Fisica  Nucleare,  Sezione  di  Catania,  95123  Catania,  Italy}
\author{R.S.~Beminiwattha}\affiliation{Louisiana Tech University, Ruston, Louisiana 71272, USA}
% ...

\date{\today}

\begin{abstract}
We report a precision measurement of the parity-violating asymmetry 
$\APV$ in the elastic scattering of longitudinally polarized electrons from $^{208}$Pb.
We measure 
$\APV =550\pm 16 {\rm (stat)}\pm 8\ {\rm (syst)}$ parts per billion, 
leading to an extraction of the neutral weak form factor 
$F_W(Q^2 = 0.00616\ {\rm GeV}^2) = 0.368 \pm 0.013$. 
Combined with our previous measurement, the extracted
neutron skin thickness is $R_n-R_p=0.283
\pm 0.071$~fm. 
The result also yields the first significant direct measurement of the interior weak density of $^{208}$Pb: $\rho^0_W = -0.0796\pm0.0036\ {\rm (exp.)}\pm0.0013\ {\rm (theo.)}\ {\rm fm}^{-3}$ leading to 
the interior baryon density $\rho^0_b = 0.1480\pm0.0036\ {\rm (exp.)}\pm0.0013\ {\rm (theo.)}\ {\rm fm}^{-3}$.
The measurement accurately constrains the density dependence of the symmetry energy of nuclear matter near saturation density, with implications for the size and composition of neutron stars.
 
\end{abstract}

\maketitle

The equation of state (EOS) of nuclear 
matter~\cite{Novario:2020kuf,Shen:2020sec,Horowitz:2019piw,Wei:2019mdj,Thiel:2019tkm} 
underlies the structure and stability of 
atomic nuclei, the formation of the elements, whether stars collapse into neutron stars or black holes, 
and the structure of neutron stars themselves.  It is remarkable that the physics of systems that vary in size by 18 orders of magnitude are governed by the same EOS. 

Observed properties of the full range of atomic nuclei, 
characterized by a nearly constant central density, 
provides critical input to the EOS which is in turn applied to infer the properties of neutron stars, first discovered by Jocelyn 
Bell Burnell~\cite{Hewish:1968bj}. 
The EOS has been used to rule out the possibility that the recently observed
2.6 solar mass object is a neutron star~\cite{Abbott:2020khf, Fattoyev:2020cws}, and could be used to infer evidence of new forms of nuclear matter, such 
as the presence of a significant nonzero strangeness component in the neutron star interior~\cite{Tolos:2020aln,Fortin:2020qin}.  

Additional constraints to the EOS are obtained from detailed studies of neutron star properties (such as size, structure, and cooling).
For example, the NICER x-ray telescope
has determined a pulsar radius to better than 10\%~\cite{Riley:2019yda},
and gravitational wave data from LIGO from a neutron star merger event 
has constrained neutron star tidal 
deformability~\cite{Chatziioannou:2020pqz,Zhang:2020azr,Guven:2020dok,Baiotti:2019sew,Piekarewicz:2019ahf,Tsang:2019mlz,Fasano:2019zwm}.

The extensive data on atomic nuclei used by EOS models do not yet constrain one critical EOS parameter, namely $L$, 
the density dependence of the 
symmetry energy.  
Recent progress with chiral effective field theory has improved theoretical constraints on $L$~\cite{PhysRevLett.125.202702}. A promising avenue to obtain experimental constraints utilizes the strong correlation between $L$ and the neutron skin thickness in heavy nuclei $R_n-R_p$, that is the difference between the rms radii of the neutron and proton distributions.
Precise data on $R_p$ are available but numerous experimental methods to determine $R_n$ suffer from uncontrolled uncertainties due to hadron dynamics~\cite{Thiel:2019tkm}.  

A more accurately interpretable method 
is to measure the neutral weak form factor $F_W$ in elastic 
electron-$^{208}$Pb scattering,
exploiting the significantly 
larger coupling of the Z$^0$ boson to neutrons compared to protons~\cite{Donnelly:1989qs,Horowitz:1999fk} to achieve an accurate $R_n$ extraction.
Such measurements can provide insights into the dependence of 
the symmetry energy on three-nucleon interactions~\cite{Bentz:2020mdk} and its role in relativistic heavy-ion collisions~\cite{Li:2019kkh}.
Weak form factors of heavy nuclei lead to a more direct extraction of the nuclear central density, which is 
governed by multinucleon interactions~\cite{Horowitz:2020evx} and may ultimately bridge to quantum chromodynamics~\cite{Drischler:2019xuo}. Well-determined nuclear weak form factors can improve the sensitivity of
dark matter searches~\cite{Yang:2019pbx} 
and tests of  neutrino-quark neutral current couplings via measurements of coherent elastic neutrino-nuclear scattering~\cite{Akimov:2017ade}.   

A precise $F_W$ extraction can be accomplished by measuring the  parity-violating asymmetry $\APV$ in longitudinally polarized elastic electron scattering off $^{208}$Pb nuclei:
\begin{equation}
\APV=\frac{\sigma_R-\sigma_L}{\sigma_R+\sigma_L}\approx
\frac{G_FQ^2|Q_W|}{4\sqrt{2}\, \pi \alpha Z}\frac{F_{\rm W}(Q^2)}{F_{\rm ch}(Q^2)},
\label{eq:APV}
\end{equation}
where $\sigma_L(\sigma_R)$ is the cross section for the scattering of left(right) handed electrons
from $^{208}$Pb, $G_F$ is the Fermi coupling constant, $F_{\rm ch}$ is the charge form factor~\cite{deForest:1966}, and $Q_W$ is the weak charge of $^{208}$Pb.  
The practical application of this 
formula requires the inclusion of Coulomb distortions~\cite{Horowitz:1998vv} and experimental parameter optimization such that a single kinematic point yields a precise $R_n$ determination~\cite{Horowitz:1999fk}.  The first measurement of $\Ameas$ for $^{208}$Pb was
published in 2012~\cite{Abrahamyan:2012gp} (PREX-1); here we report a new result (PREX-2) with greatly improved precision.

The measurement technique~\cite{Aniol:2004hp}
is driven by the requirement to measure a small asymmetry, and consequently the need to measure a high scattered electron flux. 
At the optimized kinematic point, $\APV$ is on the level of half a part per million. 
Elastically scattered electrons are isolated by a magnetic spectrometer and the high (multi-GHz) rates are measured through analog integration of detector signals. $\Ameas$ is the fractional change in detected signal between right- and left-handed electrons, repeatedly measured in short time periods using a rapid helicity flip.

The data measuring $\Ameas$ totaled  114 Coulombs of charge from a 953~MeV electron beam on a diamond-lead-diamond sandwich target at an average current of 70~$\mu$A in experimental Hall A~\cite{Alcorn:2004sb} at Thomas Jefferson National Accelerator Facility (JLab). 
The average thicknesses of the diamond and lead foils, each known to better than 5\%\ accuracy, were 90~mg/cm$^2$ and  625~mg/cm$^2$ respectively.
The scattered electrons that passed the acceptance-defining collimator at the entrance 
of each
High Resolution Spectrometer (HRS)~\cite{Alcorn:2004sb} were momentum analyzed 
and focused by three magnetic quadrupoles and a dipole.
Both the left and right HRS were equipped with identical detector packages and were positioned at their most forward angle $\approx$12.5$^{\circ}$. 
A septum magnet pair extended the reach of the spectrometers to the average desired laboratory scattering angle of $\approx$5$^{\circ}$. 
The spectrometer achieved a momentum resolution of 0.6~MeV, ensuring that the detector intercepted only elastic events; the closest inelastic state at 2.6~MeV was $\approx$0.5~MeV from the detector edge. 
The independent measurements in the left and right HRS were combined with equal statistical weight.

Individual asymmetries are  formed from 33~ms quartet or octet sequences of beam helicity, depending on the frequency of helicity reversal (either 120 or 240 Hz) created by a Pockels cell (PC)~\cite{Sinclair:2007ez} in the polarized source.
The first helicity sign in the sequence was chosen pseudorandomly, with the rest determined to form either a 
$+--+$ or $+--+-++-$ flip sequence or its complement, ensuring cancellation of 60~Hz power line noise. 
A blinding offset was added to each sequence asymmetry during decoding and maintained throughout the analysis. 
The dataset contained a little over 50 million such sequences. 

Approximately every eight hours, a half-wave plate (HWP) in the injector laser setup was toggled IN or OUT, facilitating a complete asymmetry sign reversal with no other change. The data taken between each such reversal were combined into ``slugs." Furthermore, spin manipulation in the injector beam line (using the ``double-Wien"~\cite{Sinclair:2007ez}) was changed twice during the run to add a $180^{\circ}$ precession, thereby flipping the measured asymmetry sign. With approximately equal amounts of data at each HWP/Wien state combination, these slow reversals provided critical additional cancellation of potential sources of spurious asymmetries. 

The scattering angle was calibrated using the difference in nuclear recoil between scattering from hydrogen and heavier nuclei in a water target, with tracks measured using  the vertical drift chambers in the HRS~\cite{Alcorn:2004sb}. 
The rate-averaged scattering angle was determined to be $4.71\pm0.02^\circ$ and $4.67\pm0.02^\circ$
for the left and right HRS respectively, with a four-momentum transfer squared, averaged over the combined acceptance, of  
$\langle Q^2\rangle = 0.00616 \pm 0.00005$~GeV$^2$.

The beam current was monitored with three radio frequency (rf) cavity beam current monitors (BCMs). 
The integrated charge asymmetry between positive and negative helicity bunches was determined every 7.5~seconds, and fed back to a control system which used the injector PC to minimize this quantity.
The cumulative charge correction was $20.7 \pm 0.2 $~parts per billion (ppb).
This was cross-checked to be consistent among the multiple BCMs, with a sensitivity significantly better than the ultimate $\Ameas$ statistical uncertainty. The beam trajectory throughout the accelerator complex was monitored using rf beam position monitors. Careful configuration of the polarized electron source ensured that the helicity-correlated difference in the electron beam trajectory was small:  $\approx$1~nm in beam position and $\approx$1~ppb in beam energy averaged over the entire dataset. 

The scattered electrons were detected by two identical thin fused-silica tiles ($16\times3.5\times0.5$~cm$^3$) in each spectrometer. 
With the long side of each tile oriented along the dispersive direction, approximately 7~cm was used to sample the elastically scattered electrons. The rest of the tile was a light guide to the photomultiplier tube (PMT) on the high-energy side of the elastic peak and contributed negligible background rate.
The large scattered flux ($\approx$2.2~GHz per arm) made it impractical to count individual pulses; the integrated PMT response over each helicity period provided an adequate relative measure.
The PMT and beam monitor signals were integrated and digitized by
18-bit sampling ADCs originally built for the Qweak experiment \cite{Qweak:NIM}.

The effects of beam trajectory and energy fluctuations on the detected flux were calibrated and checked using two techniques: regression over the intrinsic jitter in the beam parameters, and a dedicated, intermittent system which employed air-core dipole magnets and an rf acceleration cavity to create 15~Hz modulations of beam trajectory or energy. The dedicated calibration system was activated several times an hour throughout the data collection period.  

Table~\ref{tab:CorrSyst} lists the necessary corrections and their systematic uncertainties to extract $\Ameas = 550$ ppb from the full dataset of 96 slugs.

\begin{table}[h]
\begin{center}
\caption{Corrections and systematic uncertainties to extract $\Ameas$ listed on the bottom row with its statistical uncertainty. }
\label{tab:CorrSyst}
\begin{tabular}{l D{,}{\ \pm\ }{-1} D{,}{\ \pm\ }{-1}}
\toprule
Correction                & \multicolumn{1}{c}{Absolute [ppb]}   & \multicolumn{1}{c}{Relative [\%]}  \\
\midrule
Beam asymmetry            & -60.4 , 3.0                          &  11.0 , 0.5  \\
Charge correction         & 20.7  , 0.2                          &  3.8  , 0.0   \\
Beam polarization         & 56.8  , 5.2                          & 10.3  , 1.0  \\
Target diamond foils      & 0.7   , 1.4                          &  0.1  , 0.3  \\
Spectrometer rescattering & 0.0   , 0.1                          &  0.0  , 0.0  \\
Inelastic contributions   & 0.0   , 0.1                          &  0.0  , 0.0  \\
Transverse asymmetry      & 0.0   , 0.3                          &  0.0  , 0.1  \\
Detector nonlinearity     & 0.0   , 2.7                          &  0.0  , 0.5  \\
Angle determination       & 0.0   , 3.5                          &  0.0  , 0.6  \\
Acceptance function       & 0.0   , 2.9                          &  0.0  , 0.5  \\ \\
Total correction          & 17.7  , 8.2                          &  3.2  , 1.5  \\ \\
$\Ameas$ {\rm and statistical error}  & 550   , 16               & 100.0 , 2.9 \\ 
\bottomrule
\end{tabular}
\end{center}
\end{table}

The beam asymmetry correction accounts for helicity-correlated fluctuations in the beam trajectory (position and angle in two transverse coordinates) and energy. A set of six beam position monitors measured the transverse coordinates at locations of varying energy dispersion. The correction was calculated using a regression analysis over all measured coordinates, constrained to be consistent with the dedicated modulation data, thus optimizing precision while accounting for instrumental correlated noise and resolution. The corrections were consistent throughout the dataset, and for the grand average, with the alternative (but less precise) methods based on only regression or direct modulation-calibrated sensitivities.

\begin{figure}
\centering
\includegraphics[width=0.45\textwidth]{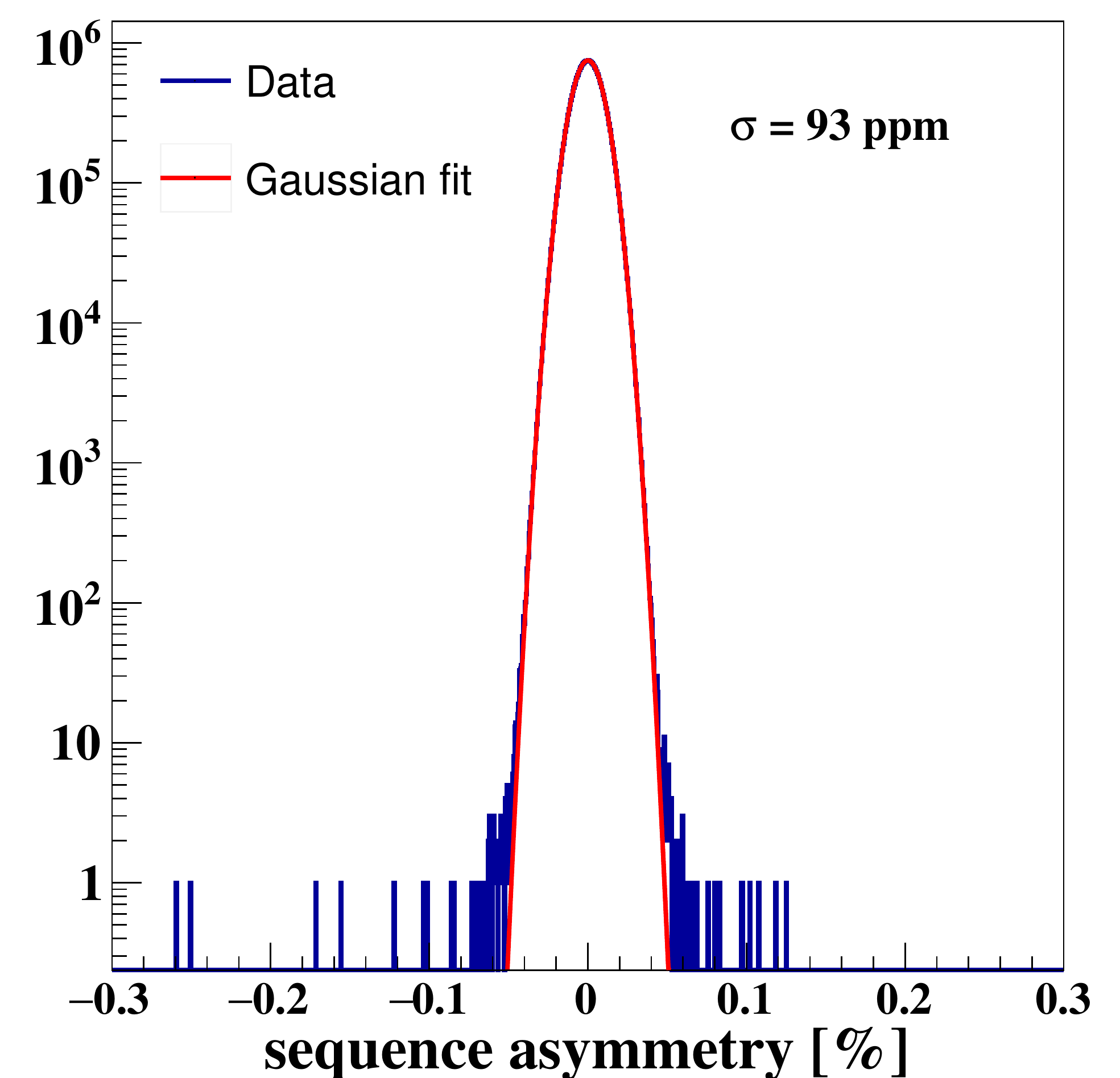}
\caption{\label{fig:apv_multiplet} Distribution of 30 million asymmetries measured over 1/30~s sequences formed with 240 Hz helicity flips. Only data taken with a beam current near to 70~$\mu$A is included.}
\end{figure}

The asymmetry data are free from any unanticipated bias as can be seen in Fig.~\ref{fig:apv_multiplet}, which shows the distribution after beam corrections of the sequence asymmetry for data collected with 240 Hz flip rate and 70~$\mu$A beam current ($\approx$62\% of the statistics). The remarkably high level of agreement between the data and the normal 
distribution fit over five orders of magnitude is achieved without the application of a single helicity-correlated data quality cut on any measured parameter.

The cumulative beam asymmetry correction was $-60.4~\pm~3.0$~ppb, where the systematic uncertainty results from assuming a 3\% uncorrelated uncertainty in the correction from each of the five beam parameters, consistent with cross-checks among various regression and beam-modulation analyses.

\begin{figure}[h]
\centering
\includegraphics[width=0.45\textwidth]{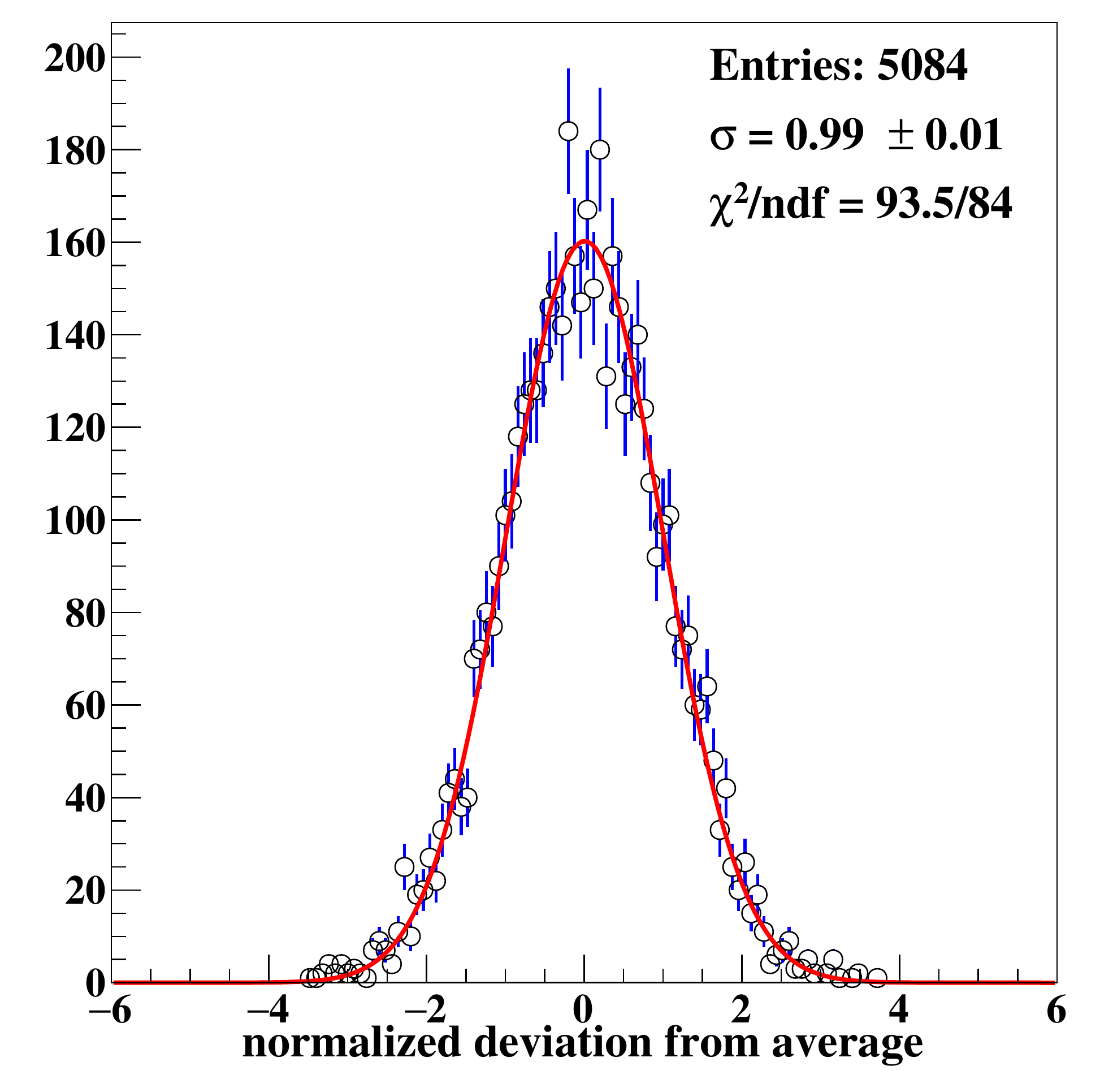}
\caption{\label{fig:minirunPull} Distribution of normalized deviations from the average (blue) for $\approx$5-minute asymmetry datasets after beam corrections, compared to a Gaussian fit(red).}
\end{figure}

The beam-corrected asymmetry data are dominated by statistical fluctuations around a single mean, as demonstrated in Fig.~\ref{fig:minirunPull}. This plot shows the deviations from the grand average asymmetry for all 5084 $\approx$5-minute data segments, with each entry normalized to its own statistical uncertainty of $\approx$1~ppm.
The data describe a normal distribution with unit variance and zero mean, as expected.

The beam-corrected asymmetry $A_{\rm corr}$ must be further corrected for the beam polarization ($P_b$),  and the background dilutions ($f_i$) and asymmetries ($A_i$) to obtain $\Ameas$:
\begin{equation}
    \Ameas = \frac{1}{P_b} \frac{A_{\rm corr} - P_b\sum_i A_i f_i}{1-\sum_i f_i}.
\end{equation}

The degree of longitudinal polarization $P_b$ of the electron beam was maximized at the beginning of data taking using the injector Mott polarimeter~\cite{Grames:2020asy}. It was periodically measured just in front of the target using a M\o ller polarimeter \cite{Alcorn:2004sb,Aulenbacher:2018weg} in dedicated low current runs that were interspersed throughout the data taking period. 
The average beam polarization result was $(89.7 \pm 0.8)\%$. The determination of the polarimeter target foil polarization was the largest contribution to the uncertainty (0.6\%).

The main background corrections are also listed in Table \ref{tab:CorrSyst}. The largest dilution ($f_C = 6.3\pm 0.5$\%) was due to the diamond foils, though the correction was small: $\APV$ for $^{12}$C and $^{208}$Pb are numerically similar. The effect of a tiny amount of scattering from magnetized pole tips in the spectrometer was found to be negligible. A 
0.26~ppb
systematic uncertainty accounted for a possible imperfect cancellation from a residual transverse electron beam polarization component; no correction was applied.

The linear response of the integrated detector signal was demonstrated to be better than 0.5\% in a bench test using a calibration system with multiple 
light sources. 
The linearity of the detector response was also monitored throughout the data taking period by comparison with BCM measurements of beam current fluctuations. 
The resulting systematic uncertainty was $2.7$~ppb; no correction was applied.

As a final sensitive test for unknown systematic effects, the data were separated into four time periods depending on the sign of the HWP and double-Wien states.  The results are statistically consistent, as summarized in Table~\ref{tab:hwpW-combo}. The $\chi^2$ for averaging over the slugs in each configuration is shown.

\begin{table}[h]
\begin{center}
\caption{$\Ameas$ for different HWP-Wien state combinations.}
\label{tab:hwpW-combo}
\begin{tabular}{lcccc}
\toprule
HWP/Wien           & $A_{\rm corr}$ sign    & $\Ameas$ [ppb]                 & $\chi^2$ & \#slugs\\ \midrule
IN / Left          &     $-$            & 540.7   $\pm$ 29.9             & 46.9     &27         \\
OUT / Left         &     $+$            & 598.8   $\pm$ 29.1             & 31.6     &29          \\
IN / Right         &     $+$            & 506.2   $\pm$ 34.1             & 18.3     &19          \\
OUT / Right        &     $-$            & 536.4   $\pm$ 37.7             & 16.0     &21          \\
\bottomrule
\end{tabular}
\end{center}
\end{table}

For a direct comparison of the measurement to theoretical predictions one must convolve the predicted asymmetry variation with the acceptance of the spectrometers:
\begin{equation}
     \langle A_{PV} \rangle = \frac{\int {\rm d}\theta \sin\theta A(\theta) \frac{{\rm d}\sigma}{{\rm d}\Omega} \epsilon(\theta)}{\int {\rm d}\theta \sin\theta\frac{{\rm d}\sigma}{{\rm d}\Omega} \epsilon(\theta)}, 
\end{equation}
where $\frac{{\rm d}\sigma}{{\rm d}\Omega}$ is the differential cross section and $A(\theta)$ is the modeled parity violating asymmetry as a function of scattering angle. The acceptance function $\epsilon(\theta)$  is defined as the relative probability for an elastically scattered electron to make it to the detector~\cite{supplemental}. The systematic uncertainty in $\epsilon(\theta)$ was determined using a simulation that took into account initial and final state radiation and multiple scattering. 

Our final results for $\Ameas$ and $F_W$ with the acceptance described by $\epsilon(\theta)$ and $\langle Q^2\rangle = 0.00616{\rm\ GeV^2}$ are:

\begin{equation*}
\begin{split}
    \Ameas & = 550 \pm 16{\rm\ (stat.)} \pm 8{\rm\ (syst.)\ ppb} \\
    F_W(\langle Q^2\rangle) & = 0.368\pm0.013{\rm\ (exp.)}\pm0.001{\rm\ (theo.)}.
\end{split}
\end{equation*}
where the experimental uncertainty in $F_W$ includes both statistical and systematic contributions.

\begin{figure}[h]
\centering
\includegraphics[width=0.45\textwidth]{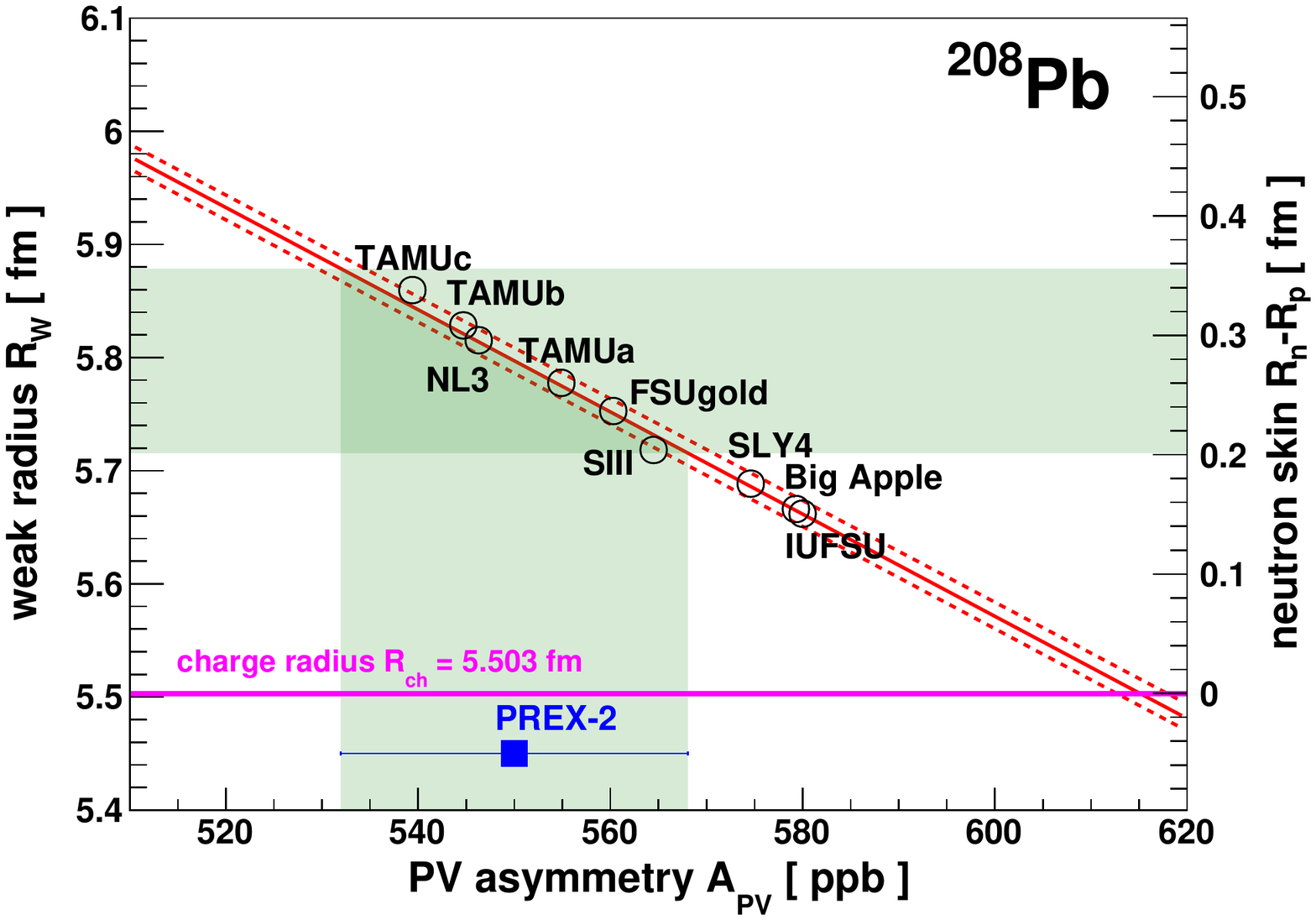}
\caption{\label{fig:apv_vs_rnp} Extraction of the weak radius (left vertical axis) or neutron skin (right vertical axis) for the $^{208}$Pb nucleus. $R_{\rm ch}$~\cite{DEVRIES1987495} is shown for comparison. }
\end{figure}

The correlation between $\APV$ and the $^{208}$Pb weak radius $R_W$ is obtained by plotting the predictions for these two quantities from a sampling of theoretical calculations~\cite{Fattoyev:2020cws,Beiner:1974gc,ToddRutel:2005zz,Fattoyev:2010mx,Lalazissis:1996rd,Fattoyev:2013yaa,Chabanat:1997un},
as shown in Fig.~\ref{fig:apv_vs_rnp}, along with the green band highlighting $\Ameas$ and its 1-$\sigma$ experimental uncertainty.

Single nucleon weak form factors are folded with point nucleon radial densities to arrive at the weak density distribution $\rho_W(r)$, using $Q_W = -117.9\pm 0.3$ which incorporates  one-loop radiative corrections 
including $\gamma$-$Z$ box contributions
\cite{Erler:2013xha,Gorchtein:2008px,Gorchtein:2011mz,nuclearGZ} as an overall constraint. 
The correlation slope in Fig.~\ref{fig:apv_vs_rnp} is determined by fitting $\rho_W(r)$ as a two-parameter Fermi function over a large variety of relativistic and nonrelativistic density functional models, determining for each model a size consistent with $R_W$ and a surface thickness $a$. 
This also determines the small model uncertainty, shown in Fig.~\ref{fig:apv_vs_rnp} (dashed red lines), corresponding to the range of  $a$~\cite{Reed:2020fdf,Horowitz:2020evx,supplemental}.

Projecting to the model correlation to determine the weak radius or alternatively the neutron skin (left and right vertical axes respectively), the PREX-2 results are 
\begin{equation*}
\begin{split}
R_W = 5.795\pm 0.082{\rm\ (exp.)} \pm 0.013{\rm\ (theo.)} ~\mbox{fm } \\
R_n-R_p = 0.278\pm 0.078{\rm\ (exp.)}\pm 0.012{\rm\ (theo.)}~\mbox{fm.}
\end{split}
\end{equation*}

The normalization constant in the Fermi-function form of $\rho_W(r)$ used to extract $R_W$ is a measure of the $^{208}$Pb interior weak density~\cite{supplemental}: 
\begin{equation*}
\rho^0_W = -0.0798\pm 0.0038{\rm\ (exp.)}\pm 0.0013{\rm\ (theo.)~\rm{fm}^{-3}}.
\end{equation*}
Combined with the well-measured 
interior charge density, the interior baryon density determined solely from the PREX-2 data is $\rho^0_b = 0.1482\pm 0.0040\ {\rm fm ^{-3}}$ (combining experimental and theoretical uncertainties).

This result is consistent with the results from the PREX-1 measurement, which found $R_n-R_p = 0.30\pm0.18$~fm~\cite{Horowitz:2012tj}. 
Table~\ref{tab:results} summarizes nuclear properties of $^{208}$Pb from the combined PREX-1 and PREX-2 results, including a 4~$\sigma$ determination of the neutron skin.

\begin{table}[h]
\begin{center}
\caption{PREX-1 and -2 combined experimental results for $^{208}$Pb. Uncertainties include both experimental and theoretical contributions.}
\label{tab:results}
\begin{tabular}{l D{,}{\ \pm\ }{-1}}
\toprule
$^{208}$Pb Parameter                & \multicolumn{1}{c}{Value}   \\ 
\midrule
Weak radius ($R_W$)                       & 5.800   ,0.075{\rm\ fm} \\
Interior weak density ($\rho^0_W$)          & -0.0796 ,0.0038{\rm\ fm^{-3}}\\
Interior baryon density ($\rho^0_b$)        & 0.1480  ,0.0038{\rm\ fm^{-3}} \\
Neutron skin ($R_n - R_p$)                & 0.283   ,0.071{\rm\ fm} \\
\bottomrule
\end{tabular}
\end{center}
\end{table}

\begin{figure}[h]
\centering
\includegraphics[width=0.47\textwidth]{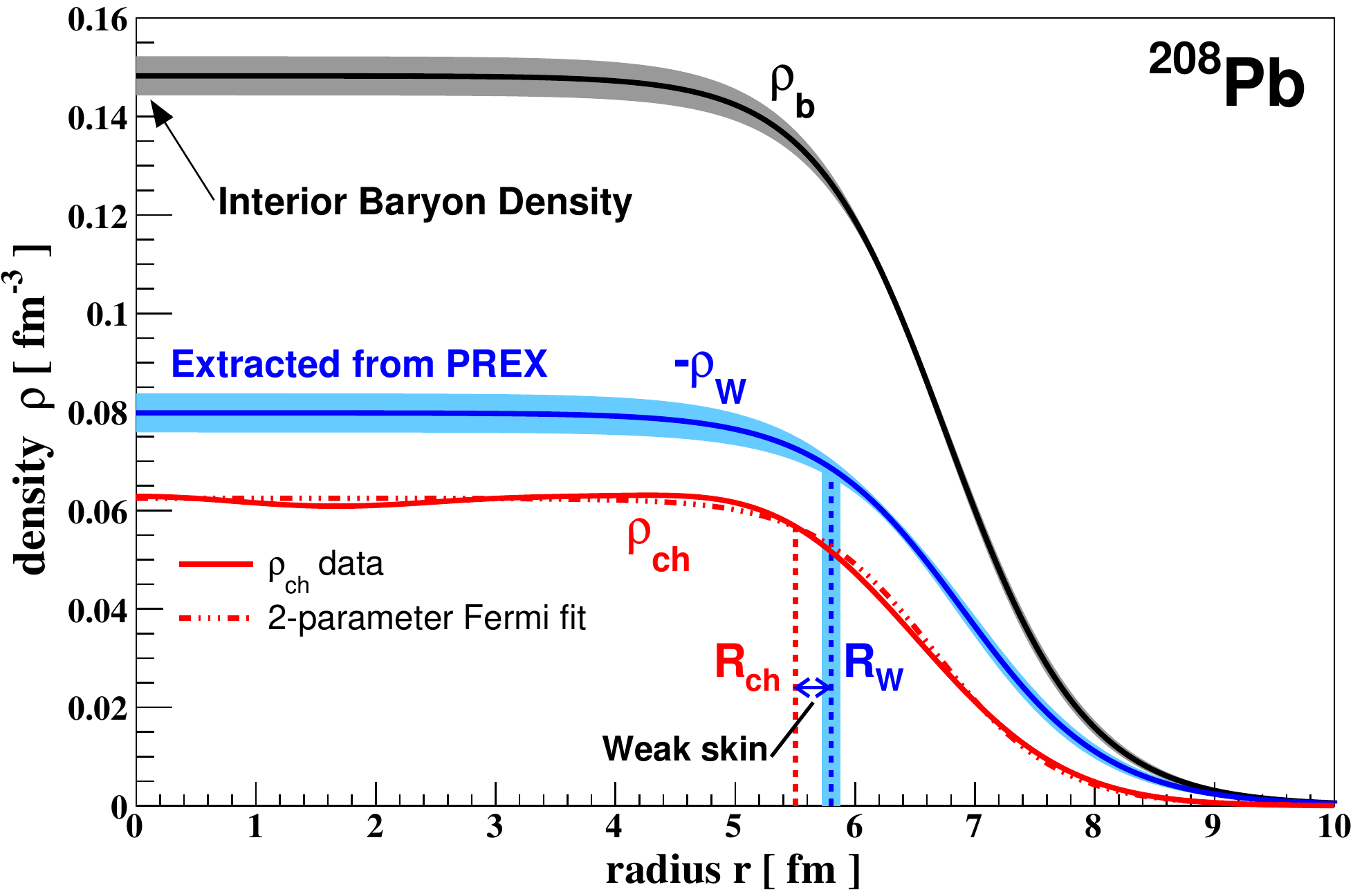}
\caption{\label{fig:centralDensity} $^{208}$Pb weak and baryon densities from the combined PREX datasets, with uncertainties shaded. The charge density~\cite{DEVRIES1987495} is also shown.}
\end{figure}
Exploiting the strong correlation between $R_n-R_p$ and the density dependence of the symmetry energy $L$, the PREX result implies a stiff symmetry energy ($L = 106 \pm 37~$MeV~\cite{Reed:2021nqk}), with important implications for critical neutron star observables. Figure~\ref{fig:centralDensity} shows the inferred radial dependence of the $^{208}$Pb charge, weak and total baryon densities together with their uncertainty bands. The precise 2.5\% determination of 
$\rho^0_b$ for $^{208}$Pb
will facilitate a sensitive examination of its close relationship to the nuclear saturation density~\cite{Horowitz:2020evx}. 

After the $^{208}$Pb run, data were also collected to measure $\Ameas$ for $^{48}$Ca (CREX)~\cite{crex-proposal}. The improved systematic control of helicity correlated beam asymmetries and several other PREX experimental innovations will inform the design of future  projects MOLLER~\cite{Benesch:2014bas}
and SoLID~\cite{solid-PVDIS} at JLab measuring fundamental electroweak couplings, as well as a more precise $^{208}$Pb radius experimental proposal at Mainz~\cite{Thiel:2019tkm,Becker:2018ggl}.

\begin{acknowledgments}
We thank the entire staff of JLab for their efforts to develop and maintain the polarized beam and the experimental apparatus, and acknowledge the support of the U.S. Department of Energy, the National Science Foundation and NSERC (Canada). This material is based upon the work supported by the U.S. Department of Energy, Office of Science, Office of Nuclear Physics Contract No. DE-AC05-06OR23177. 
\end{acknowledgments}

\bibliography{PREX2}

\providecommand{\noopsort}[1]{}\providecommand{\singleletter}[1]{#1}%
\begin{thebibliography}{57}%
\makeatletter
\providecommand \@ifxundefined [1]{%
 \@ifx{#1\undefined}
}%
\providecommand \@ifnum [1]{%
 \ifnum #1\expandafter \@firstoftwo
 \else \expandafter \@secondoftwo
 \fi
}%
\providecommand \@ifx [1]{%
 \ifx #1\expandafter \@firstoftwo
 \else \expandafter \@secondoftwo
 \fi
}%
\providecommand \natexlab [1]{#1}%
\providecommand \enquote  [1]{``#1''}%
\providecommand \bibnamefont  [1]{#1}%
\providecommand \bibfnamefont [1]{#1}%
\providecommand \citenamefont [1]{#1}%
\providecommand \href@noop [0]{\@secondoftwo}%
\providecommand \href [0]{\begingroup \@sanitize@url \@href}%
\providecommand \@href[1]{\@@startlink{#1}\@@href}%
\providecommand \@@href[1]{\endgroup#1\@@endlink}%
\providecommand \@sanitize@url [0]{\catcode `\\12\catcode `\$12\catcode
  `\&12\catcode `\#12\catcode `\^12\catcode `\_12\catcode `\%12\relax}%
\providecommand \@@startlink[1]{}%
\providecommand \@@endlink[0]{}%
\providecommand \url  [0]{\begingroup\@sanitize@url \@url }%
\providecommand \@url [1]{\endgroup\@href {#1}{\urlprefix }}%
\providecommand \urlprefix  [0]{URL }%
\providecommand \Eprint [0]{\href }%
\providecommand \doibase [0]{https://doi.org/}%
\providecommand \selectlanguage [0]{\@gobble}%
\providecommand \bibinfo  [0]{\@secondoftwo}%
\providecommand \bibfield  [0]{\@secondoftwo}%
\providecommand \translation [1]{[#1]}%
\providecommand \BibitemOpen [0]{}%
\providecommand \bibitemStop [0]{}%
\providecommand \bibitemNoStop [0]{.\EOS\space}%
\providecommand \EOS [0]{\spacefactor3000\relax}%
\providecommand \BibitemShut  [1]{\csname bibitem#1\endcsname}%
\let\auto@bib@innerbib\@empty
\bibitem [{\citenamefont {Novario}\ \emph {et~al.}(2020)\citenamefont
  {Novario}, \citenamefont {Hagen}, \citenamefont {Jansen},\ and\ \citenamefont
  {Papenbrock}}]{Novario:2020kuf}%
  \BibitemOpen
  \bibfield  {author} {\bibinfo {author} {\bibfnamefont {S.}~\bibnamefont
  {Novario}}, \bibinfo {author} {\bibfnamefont {G.}~\bibnamefont {Hagen}},
  \bibinfo {author} {\bibfnamefont {G.}~\bibnamefont {Jansen}},\ and\ \bibinfo
  {author} {\bibfnamefont {T.}~\bibnamefont {Papenbrock}},\ }\bibfield  {title}
  {\bibinfo {title} {{Charge radii of exotic neon and magnesium isotopes}},\
  }\href {https://doi.org/10.1103/PhysRevC.102.051303} {\bibfield  {journal}
  {\bibinfo  {journal} {Phys. Rev. C}\ }\textbf {\bibinfo {volume} {102}},\
  \bibinfo {pages} {051303} (\bibinfo {year} {2020})},\ \Eprint
  {https://arxiv.org/abs/2007.06684} {arXiv:2007.06684 [nucl-th]} \BibitemShut
  {NoStop}%
\bibitem [{\citenamefont {Shen}\ \emph {et~al.}(2020)\citenamefont {Shen},
  \citenamefont {Ji}, \citenamefont {Hu},\ and\ \citenamefont
  {Sumiyoshi}}]{Shen:2020sec}%
  \BibitemOpen
  \bibfield  {author} {\bibinfo {author} {\bibfnamefont {H.}~\bibnamefont
  {Shen}}, \bibinfo {author} {\bibfnamefont {F.}~\bibnamefont {Ji}}, \bibinfo
  {author} {\bibfnamefont {J.}~\bibnamefont {Hu}},\ and\ \bibinfo {author}
  {\bibfnamefont {K.}~\bibnamefont {Sumiyoshi}},\ }\bibfield  {title} {\bibinfo
  {title} {{Effects of symmetry energy on equation of state for simulations of
  core-collapse supernovae and neutron-star mergers}},\ }\href
  {https://doi.org/10.3847/1538-4357/ab72fd} {\bibfield  {journal} {\bibinfo
  {journal} {Astrophys. J.}\ }\textbf {\bibinfo {volume} {891}},\ \bibinfo
  {pages} {148} (\bibinfo {year} {2020})},\ \Eprint
  {https://arxiv.org/abs/2001.10143} {arXiv:2001.10143 [nucl-th]} \BibitemShut
  {NoStop}%
\bibitem [{\citenamefont {Horowitz}(2019)}]{Horowitz:2019piw}%
  \BibitemOpen
  \bibfield  {author} {\bibinfo {author} {\bibfnamefont {C.}~\bibnamefont
  {Horowitz}},\ }\bibfield  {title} {\bibinfo {title} {{Neutron rich matter in
  the laboratory and in the heavens after GW170817}},\ }\href
  {https://doi.org/10.1016/j.aop.2019.167992} {\bibfield  {journal} {\bibinfo
  {journal} {Ann. Phys. (Amsterdam)}\ }\textbf {\bibinfo {volume} {411}},\
  \bibinfo {pages} {167992} (\bibinfo {year} {2019})},\ \Eprint
  {https://arxiv.org/abs/1911.00411} {arXiv:1911.00411 [astro-ph.HE]}
  \BibitemShut {NoStop}%
\bibitem [{\citenamefont {Wei}\ \emph {et~al.}(2020)\citenamefont {Wei},
  \citenamefont {Lu}, \citenamefont {Burgio}, \citenamefont {Li},\ and\
  \citenamefont {Schulze}}]{Wei:2019mdj}%
  \BibitemOpen
  \bibfield  {author} {\bibinfo {author} {\bibfnamefont {J.-B.}\ \bibnamefont
  {Wei}}, \bibinfo {author} {\bibfnamefont {J.-J.}\ \bibnamefont {Lu}},
  \bibinfo {author} {\bibfnamefont {G.}~\bibnamefont {Burgio}}, \bibinfo
  {author} {\bibfnamefont {Z.}~\bibnamefont {Li}},\ and\ \bibinfo {author}
  {\bibfnamefont {H.-J.}\ \bibnamefont {Schulze}},\ }\bibfield  {title}
  {\bibinfo {title} {{Are nuclear matter properties correlated to neutron star
  observables?}},\ }\href {https://doi.org/10.1140/epja/s10050-020-00058-3}
  {\bibfield  {journal} {\bibinfo  {journal} {Eur. Phys. J. A}\ }\textbf
  {\bibinfo {volume} {56}},\ \bibinfo {pages} {63} (\bibinfo {year} {2020})},\
  \Eprint {https://arxiv.org/abs/1907.08761} {arXiv:1907.08761 [nucl-th]}
  \BibitemShut {NoStop}%
\bibitem [{\citenamefont {Thiel}\ \emph {et~al.}(2019)\citenamefont {Thiel},
  \citenamefont {Sfienti}, \citenamefont {Piekarewicz}, \citenamefont
  {Horowitz},\ and\ \citenamefont {Vanderhaeghen}}]{Thiel:2019tkm}%
  \BibitemOpen
  \bibfield  {author} {\bibinfo {author} {\bibfnamefont {M.}~\bibnamefont
  {Thiel}}, \bibinfo {author} {\bibfnamefont {C.}~\bibnamefont {Sfienti}},
  \bibinfo {author} {\bibfnamefont {J.}~\bibnamefont {Piekarewicz}}, \bibinfo
  {author} {\bibfnamefont {C.}~\bibnamefont {Horowitz}},\ and\ \bibinfo
  {author} {\bibfnamefont {M.}~\bibnamefont {Vanderhaeghen}},\ }\bibfield
  {title} {\bibinfo {title} {{Neutron skins of atomic nuclei: Per aspera ad
  astra}},\ }\href {https://doi.org/10.1088/1361-6471/ab2c6d} {\bibfield
  {journal} {\bibinfo  {journal} {J. Phys. G}\ }\textbf {\bibinfo {volume}
  {46}},\ \bibinfo {pages} {093003} (\bibinfo {year} {2019})},\ \Eprint
  {https://arxiv.org/abs/1904.12269} {arXiv:1904.12269 [nucl-ex]} \BibitemShut
  {NoStop}%
\bibitem [{\citenamefont {Hewish}\ \emph {et~al.}(1968)\citenamefont {Hewish},
  \citenamefont {Bell}, \citenamefont {Pilkington}, \citenamefont {Scott},\
  and\ \citenamefont {Collins}}]{Hewish:1968bj}%
  \BibitemOpen
  \bibfield  {author} {\bibinfo {author} {\bibfnamefont {A.}~\bibnamefont
  {Hewish}}, \bibinfo {author} {\bibfnamefont {S.}~\bibnamefont {Bell}},
  \bibinfo {author} {\bibfnamefont {J.}~\bibnamefont {Pilkington}}, \bibinfo
  {author} {\bibfnamefont {P.}~\bibnamefont {Scott}},\ and\ \bibinfo {author}
  {\bibfnamefont {R.}~\bibnamefont {Collins}},\ }\bibfield  {title} {\bibinfo
  {title} {{Observation of a rapidly pulsating radio source}},\ }\href
  {https://doi.org/10.1038/217709a0} {\bibfield  {journal} {\bibinfo  {journal}
  {Nature (London)}\ }\textbf {\bibinfo {volume} {217}},\ \bibinfo {pages}
  {709} (\bibinfo {year} {1968})}\BibitemShut {NoStop}%
\bibitem [{\citenamefont {Abbott}\ \emph {et~al.}(2020)\citenamefont {Abbott}
  \emph {et~al.}}]{Abbott:2020khf}%
  \BibitemOpen
  \bibfield  {author} {\bibinfo {author} {\bibfnamefont {R.}~\bibnamefont
  {Abbott}} \emph {et~al.} (\bibinfo {collaboration} {LIGO Scientific and Virgo
  Collaborations}),\ }\bibfield  {title} {\bibinfo {title} {{GW190814:
  Gravitational waves from the coalescence of a 23 solar mass black hole with a
  2.6 solar mass compact object}},\ }\href
  {https://doi.org/10.3847/2041-8213/ab960f} {\bibfield  {journal} {\bibinfo
  {journal} {Astrophys. J. Lett.}\ }\textbf {\bibinfo {volume} {896}},\
  \bibinfo {pages} {L44} (\bibinfo {year} {2020})},\ \Eprint
  {https://arxiv.org/abs/2006.12611} {arXiv:2006.12611 [astro-ph.HE]}
  \BibitemShut {NoStop}%
\bibitem [{\citenamefont {Fattoyev}\ \emph {et~al.}(2020)\citenamefont
  {Fattoyev}, \citenamefont {Horowitz}, \citenamefont {Piekarewicz},\ and\
  \citenamefont {Reed}}]{Fattoyev:2020cws}%
  \BibitemOpen
  \bibfield  {author} {\bibinfo {author} {\bibfnamefont {F.}~\bibnamefont
  {Fattoyev}}, \bibinfo {author} {\bibfnamefont {C.}~\bibnamefont {Horowitz}},
  \bibinfo {author} {\bibfnamefont {J.}~\bibnamefont {Piekarewicz}},\ and\
  \bibinfo {author} {\bibfnamefont {B.}~\bibnamefont {Reed}},\ }\bibfield
  {title} {\bibinfo {title} {{GW190814: Impact of a 2.6 solar mass neutron star
  on nucleonic equations of state}},\ }\href
  {https://doi.org/10.1103/PhysRevC.102.065805} {\bibfield  {journal} {\bibinfo
   {journal} {Phys. Rev. C}\ }\textbf {\bibinfo {volume} {102}},\ \bibinfo
  {pages} {065805} (\bibinfo {year} {2020})},\ \Eprint
  {https://arxiv.org/abs/2007.03799} {arXiv:2007.03799 [nucl-th]} \BibitemShut
  {NoStop}%
\bibitem [{\citenamefont {Tolos}\ and\ \citenamefont
  {Fabbietti}(2020)}]{Tolos:2020aln}%
  \BibitemOpen
  \bibfield  {author} {\bibinfo {author} {\bibfnamefont {L.}~\bibnamefont
  {Tolos}}\ and\ \bibinfo {author} {\bibfnamefont {L.}~\bibnamefont
  {Fabbietti}},\ }\bibfield  {title} {\bibinfo {title} {{Strangeness in nuclei
  and neutron stars}},\ }\href {https://doi.org/10.1016/j.ppnp.2020.103770}
  {\bibfield  {journal} {\bibinfo  {journal} {Prog. Part. Nucl. Phys.}\
  }\textbf {\bibinfo {volume} {112}},\ \bibinfo {pages} {103770} (\bibinfo
  {year} {2020})},\ \Eprint {https://arxiv.org/abs/2002.09223}
  {arXiv:2002.09223 [nucl-ex]} \BibitemShut {NoStop}%
\bibitem [{\citenamefont {Fortin}\ \emph {et~al.}(2020)\citenamefont {Fortin},
  \citenamefont {Raduta}, \citenamefont {Avancini},\ and\ \citenamefont
  {Provid\^encia}}]{Fortin:2020qin}%
  \BibitemOpen
  \bibfield  {author} {\bibinfo {author} {\bibfnamefont {M.}~\bibnamefont
  {Fortin}}, \bibinfo {author} {\bibfnamefont {A.~R.}\ \bibnamefont {Raduta}},
  \bibinfo {author} {\bibfnamefont {S.}~\bibnamefont {Avancini}},\ and\
  \bibinfo {author} {\bibfnamefont {C.}~\bibnamefont {Provid\^encia}},\
  }\bibfield  {title} {\bibinfo {title} {{Relativistic hypernuclear compact
  stars with calibrated equations of state}},\ }\href
  {https://doi.org/10.1103/PhysRevD.101.034017} {\bibfield  {journal} {\bibinfo
   {journal} {Phys. Rev. D}\ }\textbf {\bibinfo {volume} {101}},\ \bibinfo
  {pages} {034017} (\bibinfo {year} {2020})},\ \Eprint
  {https://arxiv.org/abs/2001.08036} {arXiv:2001.08036 [hep-ph]} \BibitemShut
  {NoStop}%
\bibitem [{\citenamefont {Riley}\ \emph {et~al.}(2019)\citenamefont {Riley}
  \emph {et~al.}}]{Riley:2019yda}%
  \BibitemOpen
  \bibfield  {author} {\bibinfo {author} {\bibfnamefont {T.~E.}\ \bibnamefont
  {Riley}} \emph {et~al.},\ }\bibfield  {title} {\bibinfo {title} {{A $NICER$
  View of PSR J0030+0451: Millisecond pulsar parameter estimation}},\ }\href
  {https://doi.org/10.3847/2041-8213/ab481c} {\bibfield  {journal} {\bibinfo
  {journal} {Astrophys. J. Lett.}\ }\textbf {\bibinfo {volume} {887}},\
  \bibinfo {pages} {L21} (\bibinfo {year} {2019})},\ \Eprint
  {https://arxiv.org/abs/1912.05702} {arXiv:1912.05702 [astro-ph.HE]}
  \BibitemShut {NoStop}%
\bibitem [{\citenamefont {Chatziioannou}(2020)}]{Chatziioannou:2020pqz}%
  \BibitemOpen
  \bibfield  {author} {\bibinfo {author} {\bibfnamefont {K.}~\bibnamefont
  {Chatziioannou}},\ }\bibfield  {title} {\bibinfo {title} {{Neutron star tidal
  deformability and equation of state constraints}},\ }\href
  {https://doi.org/10.1007/s10714-020-02754-3} {\bibfield  {journal} {\bibinfo
  {journal} {Gen. Relativ. Gravit.}\ }\textbf {\bibinfo {volume} {52}},\
  \bibinfo {pages} {109} (\bibinfo {year} {2020})},\ \Eprint
  {https://arxiv.org/abs/2006.03168} {arXiv:2006.03168 [gr-qc]} \BibitemShut
  {NoStop}%
\bibitem [{\citenamefont {Zhang}\ \emph {et~al.}(2020)\citenamefont {Zhang},
  \citenamefont {Liu}, \citenamefont {Xia}, \citenamefont {Li},\ and\
  \citenamefont {Biswal}}]{Zhang:2020azr}%
  \BibitemOpen
  \bibfield  {author} {\bibinfo {author} {\bibfnamefont {Y.}~\bibnamefont
  {Zhang}}, \bibinfo {author} {\bibfnamefont {M.}~\bibnamefont {Liu}}, \bibinfo
  {author} {\bibfnamefont {C.-J.}\ \bibnamefont {Xia}}, \bibinfo {author}
  {\bibfnamefont {Z.}~\bibnamefont {Li}},\ and\ \bibinfo {author}
  {\bibfnamefont {S.~K.}\ \bibnamefont {Biswal}},\ }\bibfield  {title}
  {\bibinfo {title} {{Constraints on the symmetry energy and its associated
  parameters from nuclei to neutron stars}},\ }\href
  {https://doi.org/10.1103/PhysRevC.101.034303} {\bibfield  {journal} {\bibinfo
   {journal} {Phys. Rev. C}\ }\textbf {\bibinfo {volume} {101}},\ \bibinfo
  {pages} {034303} (\bibinfo {year} {2020})},\ \Eprint
  {https://arxiv.org/abs/2002.10884} {arXiv:2002.10884 [nucl-th]} \BibitemShut
  {NoStop}%
\bibitem [{\citenamefont {G\"uven}\ \emph {et~al.}(2020)\citenamefont
  {G\"uven}, \citenamefont {Bozkurt}, \citenamefont {Khan},\ and\ \citenamefont
  {Margueron}}]{Guven:2020dok}%
  \BibitemOpen
  \bibfield  {author} {\bibinfo {author} {\bibfnamefont {H.}~\bibnamefont
  {G\"uven}}, \bibinfo {author} {\bibfnamefont {K.}~\bibnamefont {Bozkurt}},
  \bibinfo {author} {\bibfnamefont {E.}~\bibnamefont {Khan}},\ and\ \bibinfo
  {author} {\bibfnamefont {J.}~\bibnamefont {Margueron}},\ }\bibfield  {title}
  {\bibinfo {title} {{Multimessenger and multiphysics Bayesian inference for
  the GW170817 binary neutron star merger}},\ }\href
  {https://doi.org/10.1103/PhysRevC.102.015805} {\bibfield  {journal} {\bibinfo
   {journal} {Phys. Rev. C}\ }\textbf {\bibinfo {volume} {102}},\ \bibinfo
  {pages} {015805} (\bibinfo {year} {2020})},\ \Eprint
  {https://arxiv.org/abs/2001.10259} {arXiv:2001.10259 [nucl-th]} \BibitemShut
  {NoStop}%
\bibitem [{\citenamefont {Baiotti}(2019)}]{Baiotti:2019sew}%
  \BibitemOpen
  \bibfield  {author} {\bibinfo {author} {\bibfnamefont {L.}~\bibnamefont
  {Baiotti}},\ }\bibfield  {title} {\bibinfo {title} {{Gravitational waves from
  neutron star mergers and their relation to the nuclear equation of state}},\
  }\href {https://doi.org/10.1016/j.ppnp.2019.103714} {\bibfield  {journal}
  {\bibinfo  {journal} {Prog. Part. Nucl. Phys.}\ }\textbf {\bibinfo {volume}
  {109}},\ \bibinfo {pages} {103714} (\bibinfo {year} {2019})},\ \Eprint
  {https://arxiv.org/abs/1907.08534} {arXiv:1907.08534 [astro-ph.HE]}
  \BibitemShut {NoStop}%
\bibitem [{\citenamefont {Piekarewicz}\ and\ \citenamefont
  {Fattoyev}(2019)}]{Piekarewicz:2019ahf}%
  \BibitemOpen
  \bibfield  {author} {\bibinfo {author} {\bibfnamefont {J.}~\bibnamefont
  {Piekarewicz}}\ and\ \bibinfo {author} {\bibfnamefont {F.}~\bibnamefont
  {Fattoyev}},\ }\bibfield  {title} {\bibinfo {title} {{Neutron rich matter in
  heaven and on Earth}},\ }\href {https://doi.org/10.1063/PT.3.4247} {\bibfield
   {journal} {\bibinfo  {journal} {Phys. Today}\ }\textbf {\bibinfo {volume}
  {7}},\ \bibinfo {pages} {30} (\bibinfo {year} {2019})},\ \Eprint
  {https://arxiv.org/abs/1907.02561} {arXiv:1907.02561 [nucl-th]} \BibitemShut
  {NoStop}%
\bibitem [{\citenamefont {Tsang}\ \emph {et~al.}(2019)\citenamefont {Tsang},
  \citenamefont {Lynch}, \citenamefont {Danielewicz},\ and\ \citenamefont
  {Tsang}}]{Tsang:2019mlz}%
  \BibitemOpen
  \bibfield  {author} {\bibinfo {author} {\bibfnamefont {M.}~\bibnamefont
  {Tsang}}, \bibinfo {author} {\bibfnamefont {W.}~\bibnamefont {Lynch}},
  \bibinfo {author} {\bibfnamefont {P.}~\bibnamefont {Danielewicz}},\ and\
  \bibinfo {author} {\bibfnamefont {C.}~\bibnamefont {Tsang}},\ }\bibfield
  {title} {\bibinfo {title} {{Symmetry energy constraints from GW170817 and
  laboratory experiments}},\ }\href
  {https://doi.org/10.1016/j.physletb.2019.06.059} {\bibfield  {journal}
  {\bibinfo  {journal} {Phys. Lett. B}\ }\textbf {\bibinfo {volume} {795}},\
  \bibinfo {pages} {533} (\bibinfo {year} {2019})},\ \Eprint
  {https://arxiv.org/abs/1906.02180} {arXiv:1906.02180 [nucl-ex]} \BibitemShut
  {NoStop}%
\bibitem [{\citenamefont {Fasano}\ \emph {et~al.}(2019)\citenamefont {Fasano},
  \citenamefont {Abdelsalhin}, \citenamefont {Maselli},\ and\ \citenamefont
  {Ferrari}}]{Fasano:2019zwm}%
  \BibitemOpen
  \bibfield  {author} {\bibinfo {author} {\bibfnamefont {M.}~\bibnamefont
  {Fasano}}, \bibinfo {author} {\bibfnamefont {T.}~\bibnamefont {Abdelsalhin}},
  \bibinfo {author} {\bibfnamefont {A.}~\bibnamefont {Maselli}},\ and\ \bibinfo
  {author} {\bibfnamefont {V.}~\bibnamefont {Ferrari}},\ }\bibfield  {title}
  {\bibinfo {title} {{Constraining the Neutron Star Equation of State Using
  Multiband Independent Measurements of Radii and Tidal Deformabilities}},\
  }\href {https://doi.org/10.1103/PhysRevLett.123.141101} {\bibfield  {journal}
  {\bibinfo  {journal} {Phys. Rev. Lett.}\ }\textbf {\bibinfo {volume} {123}},\
  \bibinfo {pages} {141101} (\bibinfo {year} {2019})},\ \Eprint
  {https://arxiv.org/abs/1902.05078} {arXiv:1902.05078 [astro-ph.HE]}
  \BibitemShut {NoStop}%
\bibitem [{\citenamefont {Drischler}\ \emph {et~al.}(2020)\citenamefont
  {Drischler}, \citenamefont {Furnstahl}, \citenamefont {Melendez},\ and\
  \citenamefont {Phillips}}]{PhysRevLett.125.202702}%
  \BibitemOpen
  \bibfield  {author} {\bibinfo {author} {\bibfnamefont {C.}~\bibnamefont
  {Drischler}}, \bibinfo {author} {\bibfnamefont {R.~J.}\ \bibnamefont
  {Furnstahl}}, \bibinfo {author} {\bibfnamefont {J.~A.}\ \bibnamefont
  {Melendez}},\ and\ \bibinfo {author} {\bibfnamefont {D.~R.}\ \bibnamefont
  {Phillips}},\ }\bibfield  {title} {\bibinfo {title} {How well do we know the
  neutron-matter equation of state at the densities inside neutron stars? a
  bayesian approach with correlated uncertainties},\ }\href
  {https://doi.org/10.1103/PhysRevLett.125.202702} {\bibfield  {journal}
  {\bibinfo  {journal} {Phys. Rev. Lett.}\ }\textbf {\bibinfo {volume} {125}},\
  \bibinfo {pages} {202702} (\bibinfo {year} {2020})}\BibitemShut {NoStop}%
\bibitem [{\citenamefont {Donnelly}\ \emph {et~al.}(1989)\citenamefont
  {Donnelly}, \citenamefont {Dubach},\ and\ \citenamefont
  {Sick}}]{Donnelly:1989qs}%
  \BibitemOpen
  \bibfield  {author} {\bibinfo {author} {\bibfnamefont {T.}~\bibnamefont
  {Donnelly}}, \bibinfo {author} {\bibfnamefont {J.}~\bibnamefont {Dubach}},\
  and\ \bibinfo {author} {\bibfnamefont {I.}~\bibnamefont {Sick}},\ }\bibfield
  {title} {\bibinfo {title} {{Isospin dependences in parity violating electron
  scattering}},\ }\href {https://doi.org/10.1016/0375-9474(89)90432-6}
  {\bibfield  {journal} {\bibinfo  {journal} {Nucl. Phys. A}\ }\textbf
  {\bibinfo {volume} {503}},\ \bibinfo {pages} {589} (\bibinfo {year}
  {1989})}\BibitemShut {NoStop}%
\bibitem [{\citenamefont {Horowitz}\ \emph {et~al.}(2001)\citenamefont
  {Horowitz}, \citenamefont {Pollock}, \citenamefont {Souder},\ and\
  \citenamefont {Michaels}}]{Horowitz:1999fk}%
  \BibitemOpen
  \bibfield  {author} {\bibinfo {author} {\bibfnamefont {C.}~\bibnamefont
  {Horowitz}}, \bibinfo {author} {\bibfnamefont {S.}~\bibnamefont {Pollock}},
  \bibinfo {author} {\bibfnamefont {P.}~\bibnamefont {Souder}},\ and\ \bibinfo
  {author} {\bibfnamefont {R.}~\bibnamefont {Michaels}},\ }\bibfield  {title}
  {\bibinfo {title} {{Parity violating measurements of neutron densities}},\
  }\href {https://doi.org/10.1103/PhysRevC.63.025501} {\bibfield  {journal}
  {\bibinfo  {journal} {Phys. Rev. C}\ }\textbf {\bibinfo {volume} {63}},\
  \bibinfo {pages} {025501} (\bibinfo {year} {2001})},\ \Eprint
  {https://arxiv.org/abs/nucl-th/9912038} {arXiv:nucl-th/9912038} \BibitemShut
  {NoStop}%
\bibitem [{\citenamefont {Bentz}\ and\ \citenamefont
  {Clo\"et}(2020)}]{Bentz:2020mdk}%
  \BibitemOpen
  \bibfield  {author} {\bibinfo {author} {\bibfnamefont {W.}~\bibnamefont
  {Bentz}}\ and\ \bibinfo {author} {\bibfnamefont {I.~C.}\ \bibnamefont
  {Clo\"et}},\ }\href@noop {} {\bibinfo {title} {{Symmetry energy of nuclear
  matter and isovector three-particle interactions}}} (\bibinfo {year}
  {2020}),\ \Eprint {https://arxiv.org/abs/2004.11605} {arXiv:2004.11605}
  \BibitemShut {NoStop}%
\bibitem [{\citenamefont {Li}\ \emph {et~al.}(2020)\citenamefont {Li},
  \citenamefont {Xu}, \citenamefont {Zhou}, \citenamefont {Wang}, \citenamefont
  {Zhao}, \citenamefont {Chen},\ and\ \citenamefont {Wang}}]{Li:2019kkh}%
  \BibitemOpen
  \bibfield  {author} {\bibinfo {author} {\bibfnamefont {H.}~\bibnamefont
  {Li}}, \bibinfo {author} {\bibfnamefont {H.-j.}\ \bibnamefont {Xu}}, \bibinfo
  {author} {\bibfnamefont {Y.}~\bibnamefont {Zhou}}, \bibinfo {author}
  {\bibfnamefont {X.}~\bibnamefont {Wang}}, \bibinfo {author} {\bibfnamefont
  {J.}~\bibnamefont {Zhao}}, \bibinfo {author} {\bibfnamefont {L.-W.}\
  \bibnamefont {Chen}},\ and\ \bibinfo {author} {\bibfnamefont
  {F.}~\bibnamefont {Wang}},\ }\bibfield  {title} {\bibinfo {title} {{Probing
  the Neutron Skin with Ultrarelativistic Isobaric Collisions}},\ }\href
  {https://doi.org/10.1103/PhysRevLett.125.222301} {\bibfield  {journal}
  {\bibinfo  {journal} {Phys. Rev. Lett.}\ }\textbf {\bibinfo {volume} {125}},\
  \bibinfo {pages} {222301} (\bibinfo {year} {2020})},\ \Eprint
  {https://arxiv.org/abs/1910.06170} {arXiv:1910.06170 [nucl-th]} \BibitemShut
  {NoStop}%
\bibitem [{\citenamefont {Horowitz}\ \emph {et~al.}(2020)\citenamefont
  {Horowitz}, \citenamefont {Piekarewicz},\ and\ \citenamefont
  {Reed}}]{Horowitz:2020evx}%
  \BibitemOpen
  \bibfield  {author} {\bibinfo {author} {\bibfnamefont {C.}~\bibnamefont
  {Horowitz}}, \bibinfo {author} {\bibfnamefont {J.}~\bibnamefont
  {Piekarewicz}},\ and\ \bibinfo {author} {\bibfnamefont {B.}~\bibnamefont
  {Reed}},\ }\bibfield  {title} {\bibinfo {title} {{Insights into nuclear
  saturation density from parity violating electron scattering}},\ }\href
  {https://doi.org/10.1103/PhysRevC.102.044321} {\bibfield  {journal} {\bibinfo
   {journal} {Phys. Rev. C}\ }\textbf {\bibinfo {volume} {102}},\ \bibinfo
  {pages} {044321} (\bibinfo {year} {2020})},\ \Eprint
  {https://arxiv.org/abs/2007.07117} {arXiv:2007.07117 [nucl-th]} \BibitemShut
  {NoStop}%
\bibitem [{\citenamefont {Drischler}\ \emph {et~al.}(2019)\citenamefont
  {Drischler}, \citenamefont {Haxton}, \citenamefont {McElvain}, \citenamefont
  {Mereghetti}, \citenamefont {Nicholson}, \citenamefont {Vranas},\ and\
  \citenamefont {Walker-Loud}}]{Drischler:2019xuo}%
  \BibitemOpen
  \bibfield  {author} {\bibinfo {author} {\bibfnamefont {C.}~\bibnamefont
  {Drischler}}, \bibinfo {author} {\bibfnamefont {W.}~\bibnamefont {Haxton}},
  \bibinfo {author} {\bibfnamefont {K.}~\bibnamefont {McElvain}}, \bibinfo
  {author} {\bibfnamefont {E.}~\bibnamefont {Mereghetti}}, \bibinfo {author}
  {\bibfnamefont {A.}~\bibnamefont {Nicholson}}, \bibinfo {author}
  {\bibfnamefont {P.}~\bibnamefont {Vranas}},\ and\ \bibinfo {author}
  {\bibfnamefont {A.}~\bibnamefont {Walker-Loud}},\ }\bibfield  {title}
  {\bibinfo {title} {{Towards grounding nuclear physics in QCD}}\ }(\bibinfo
  {year} {2019})\ \Eprint {https://arxiv.org/abs/1910.07961} {arXiv:1910.07961}
  \BibitemShut {NoStop}%
\bibitem [{\citenamefont {Yang}\ \emph {et~al.}(2019)\citenamefont {Yang},
  \citenamefont {Hernandez},\ and\ \citenamefont {Piekarewicz}}]{Yang:2019pbx}%
  \BibitemOpen
  \bibfield  {author} {\bibinfo {author} {\bibfnamefont {J.}~\bibnamefont
  {Yang}}, \bibinfo {author} {\bibfnamefont {J.~A.}\ \bibnamefont
  {Hernandez}},\ and\ \bibinfo {author} {\bibfnamefont {J.}~\bibnamefont
  {Piekarewicz}},\ }\bibfield  {title} {\bibinfo {title} {{Electroweak probes
  of ground state densities}},\ }\href
  {https://doi.org/10.1103/PhysRevC.100.054301} {\bibfield  {journal} {\bibinfo
   {journal} {Phys. Rev. C}\ }\textbf {\bibinfo {volume} {100}},\ \bibinfo
  {pages} {054301} (\bibinfo {year} {2019})},\ \Eprint
  {https://arxiv.org/abs/1908.10939} {arXiv:1908.10939 [nucl-th]} \BibitemShut
  {NoStop}%
\bibitem [{\citenamefont {Akimov}\ \emph {et~al.}(2017)\citenamefont {Akimov}
  \emph {et~al.}}]{Akimov:2017ade}%
  \BibitemOpen
  \bibfield  {author} {\bibinfo {author} {\bibfnamefont {D.}~\bibnamefont
  {Akimov}} \emph {et~al.} (\bibinfo {collaboration} {COHERENT
  Collaboration}),\ }\bibfield  {title} {\bibinfo {title} {{Observation of
  coherent elastic neutrino-nucleus scattering}},\ }\href
  {https://doi.org/10.1126/science.aao0990} {\bibfield  {journal} {\bibinfo
  {journal} {Science}\ }\textbf {\bibinfo {volume} {357}},\ \bibinfo {pages}
  {1123} (\bibinfo {year} {2017})},\ \Eprint {https://arxiv.org/abs/1708.01294}
  {arXiv:1708.01294 [nucl-ex]} \BibitemShut {NoStop}%
\bibitem [{\citenamefont {de~Forest~Jr.}\ and\ \citenamefont
  {Walecka}(1966)}]{deForest:1966}%
  \BibitemOpen
  \bibfield  {author} {\bibinfo {author} {\bibfnamefont {T.}~\bibnamefont
  {de~Forest~Jr.}}\ and\ \bibinfo {author} {\bibfnamefont {J.}~\bibnamefont
  {Walecka}},\ }\bibfield  {title} {\bibinfo {title} {Electron scattering and
  nuclear structure},\ }\href {https://doi.org/10.1080/00018736600101254}
  {\bibfield  {journal} {\bibinfo  {journal} {Adv. Phys.}\ }\textbf {\bibinfo
  {volume} {15}},\ \bibinfo {pages} {1} (\bibinfo {year} {1966})}\BibitemShut
  {NoStop}%
\bibitem [{\citenamefont {Horowitz}(1998)}]{Horowitz:1998vv}%
  \BibitemOpen
  \bibfield  {author} {\bibinfo {author} {\bibfnamefont {C.}~\bibnamefont
  {Horowitz}},\ }\bibfield  {title} {\bibinfo {title} {{Parity violating
  elastic electron scattering and Coulomb distortions}},\ }\href
  {https://doi.org/10.1103/PhysRevC.57.3430} {\bibfield  {journal} {\bibinfo
  {journal} {Phys. Rev. C}\ }\textbf {\bibinfo {volume} {57}},\ \bibinfo
  {pages} {3430} (\bibinfo {year} {1998})},\ \Eprint
  {https://arxiv.org/abs/nucl-th/9801011} {arXiv:nucl-th/9801011} \BibitemShut
  {NoStop}%
\bibitem [{\citenamefont {Abrahamyan}\ \emph {et~al.}(2012)\citenamefont
  {Abrahamyan} \emph {et~al.}}]{Abrahamyan:2012gp}%
  \BibitemOpen
  \bibfield  {author} {\bibinfo {author} {\bibfnamefont {S.}~\bibnamefont
  {Abrahamyan}} \emph {et~al.},\ }\bibfield  {title} {\bibinfo {title}
  {{Measurement of the Neutron Radius of 208Pb Through Parity-Violation in
  Electron Scattering}},\ }\href
  {https://doi.org/10.1103/PhysRevLett.108.112502} {\bibfield  {journal}
  {\bibinfo  {journal} {Phys. Rev. Lett.}\ }\textbf {\bibinfo {volume} {108}},\
  \bibinfo {pages} {112502} (\bibinfo {year} {2012})},\ \Eprint
  {https://arxiv.org/abs/1201.2568} {arXiv:1201.2568 [nucl-ex]} \BibitemShut
  {NoStop}%
\bibitem [{\citenamefont {Aniol}\ \emph {et~al.}(2004)\citenamefont {Aniol}
  \emph {et~al.}}]{Aniol:2004hp}%
  \BibitemOpen
  \bibfield  {author} {\bibinfo {author} {\bibfnamefont {K.~A.}\ \bibnamefont
  {Aniol}} \emph {et~al.} (\bibinfo {collaboration} {HAPPEX Collaboration}),\
  }\bibfield  {title} {\bibinfo {title} {{Parity violating electroweak
  asymmetry in polarized-e p scattering}},\ }\href
  {https://doi.org/10.1103/PhysRevC.69.065501} {\bibfield  {journal} {\bibinfo
  {journal} {Phys. Rev. C}\ }\textbf {\bibinfo {volume} {69}},\ \bibinfo
  {pages} {065501} (\bibinfo {year} {2004})},\ \Eprint
  {https://arxiv.org/abs/nucl-ex/0402004} {arXiv:nucl-ex/0402004} \BibitemShut
  {NoStop}%
\bibitem [{\citenamefont {Alcorn}\ \emph {et~al.}(2004)\citenamefont {Alcorn}
  \emph {et~al.}}]{Alcorn:2004sb}%
  \BibitemOpen
  \bibfield  {author} {\bibinfo {author} {\bibfnamefont {J.}~\bibnamefont
  {Alcorn}} \emph {et~al.},\ }\bibfield  {title} {\bibinfo {title} {{Basic
  instrumentation for Hall A at Jefferson Lab}},\ }\href
  {https://doi.org/10.1016/j.nima.2003.11.415} {\bibfield  {journal} {\bibinfo
  {journal} {Nucl. Instrum. Meth. A}\ }\textbf {\bibinfo {volume} {522}},\
  \bibinfo {pages} {294} (\bibinfo {year} {2004})}\BibitemShut {NoStop}%
\bibitem [{\citenamefont {Sinclair}\ \emph {et~al.}(2007)\citenamefont
  {Sinclair}, \citenamefont {Adderley}, \citenamefont {Dunham}, \citenamefont
  {Hansknecht}, \citenamefont {Hartmann}, \citenamefont {Poelker},
  \citenamefont {Price}, \citenamefont {Rutt}, \citenamefont {Schneider},\ and\
  \citenamefont {Steigerwald}}]{Sinclair:2007ez}%
  \BibitemOpen
  \bibfield  {author} {\bibinfo {author} {\bibfnamefont {C.}~\bibnamefont
  {Sinclair}}, \bibinfo {author} {\bibfnamefont {P.}~\bibnamefont {Adderley}},
  \bibinfo {author} {\bibfnamefont {B.}~\bibnamefont {Dunham}}, \bibinfo
  {author} {\bibfnamefont {J.}~\bibnamefont {Hansknecht}}, \bibinfo {author}
  {\bibfnamefont {P.}~\bibnamefont {Hartmann}}, \bibinfo {author}
  {\bibfnamefont {M.}~\bibnamefont {Poelker}}, \bibinfo {author} {\bibfnamefont
  {J.}~\bibnamefont {Price}}, \bibinfo {author} {\bibfnamefont
  {P.}~\bibnamefont {Rutt}}, \bibinfo {author} {\bibfnamefont {W.}~\bibnamefont
  {Schneider}},\ and\ \bibinfo {author} {\bibfnamefont {M.}~\bibnamefont
  {Steigerwald}},\ }\bibfield  {title} {\bibinfo {title} {{Development of a
  high average current polarized electron source with long cathode operational
  lifetime}},\ }\href {https://doi.org/10.1103/PhysRevSTAB.10.023501}
  {\bibfield  {journal} {\bibinfo  {journal} {Phys. Rev. ST Accel. Beams}\
  }\textbf {\bibinfo {volume} {10}},\ \bibinfo {pages} {023501} (\bibinfo
  {year} {2007})}\BibitemShut {NoStop}%
\bibitem [{\citenamefont {Allison}\ \emph {et~al.}(2013)\citenamefont {Allison}
  \emph {et~al.}}]{Qweak:NIM}%
  \BibitemOpen
  \bibfield  {author} {\bibinfo {author} {\bibfnamefont {T.}~\bibnamefont
  {Allison}} \emph {et~al.},\ }\bibfield  {title} {\bibinfo {title} {{The Qweak
  Experimental Apparatus}},\ }\href@noop {} {\bibfield  {journal} {\bibinfo
  {journal} {Nucl. Instrum. Meth. A}\ }\textbf {\bibinfo {volume} {781}},\
  \bibinfo {pages} {105} (\bibinfo {year} {2013})}\BibitemShut {NoStop}%
\bibitem [{\citenamefont {Grames}\ \emph {et~al.}(2020)\citenamefont {Grames}
  \emph {et~al.}}]{Grames:2020asy}%
  \BibitemOpen
  \bibfield  {author} {\bibinfo {author} {\bibfnamefont {J.~M.}\ \bibnamefont
  {Grames}} \emph {et~al.},\ }\bibfield  {title} {\bibinfo {title} {{High
  precision 5 MeV Mott polarimeter}},\ }\href
  {https://doi.org/10.1103/PhysRevC.102.015501} {\bibfield  {journal} {\bibinfo
   {journal} {Phys. Rev. C}\ }\textbf {\bibinfo {volume} {102}},\ \bibinfo
  {pages} {015501} (\bibinfo {year} {2020})}\BibitemShut {NoStop}%
\bibitem [{\citenamefont {Aulenbacher}\ \emph {et~al.}(2018)\citenamefont
  {Aulenbacher}, \citenamefont {Chudakov}, \citenamefont {Gaskell},
  \citenamefont {Grames},\ and\ \citenamefont {Paschke}}]{Aulenbacher:2018weg}%
  \BibitemOpen
  \bibfield  {author} {\bibinfo {author} {\bibfnamefont {K.}~\bibnamefont
  {Aulenbacher}}, \bibinfo {author} {\bibfnamefont {E.}~\bibnamefont
  {Chudakov}}, \bibinfo {author} {\bibfnamefont {D.}~\bibnamefont {Gaskell}},
  \bibinfo {author} {\bibfnamefont {J.}~\bibnamefont {Grames}},\ and\ \bibinfo
  {author} {\bibfnamefont {K.~D.}\ \bibnamefont {Paschke}},\ }\bibfield
  {title} {\bibinfo {title} {{Precision electron beam polarimetry for next
  generation nuclear physics experiments}},\ }\href
  {https://doi.org/10.1142/S0218301318300047} {\bibfield  {journal} {\bibinfo
  {journal} {Int. J. Mod. Phys. E}\ }\textbf {\bibinfo {volume} {27}},\
  \bibinfo {pages} {1830004} (\bibinfo {year} {2018})}\BibitemShut {NoStop}%
\bibitem [{sup()}]{supplemental}%
  \BibitemOpen
  \href@noop {} {}\bibinfo {howpublished} {See the Supplemental Material for
  the acceptance function and further details on the Fermi-function fit, which
  include Refs.~\cite{Sprung_1997,Piekarewicz:2016vbn}.}\BibitemShut {Stop}%
\bibitem [{\citenamefont {Sprung}\ and\ \citenamefont
  {Martorell}(1997)}]{Sprung_1997}%
  \BibitemOpen
  \bibfield  {author} {\bibinfo {author} {\bibfnamefont {D.~W.~L.}\
  \bibnamefont {Sprung}}\ and\ \bibinfo {author} {\bibfnamefont
  {J.}~\bibnamefont {Martorell}},\ }\bibfield  {title} {\bibinfo {title} {The
  symmetrized fermi function and its transforms},\ }\href
  {https://doi.org/10.1088/0305-4470/30/18/026} {\bibfield  {journal} {\bibinfo
   {journal} {J. Phys. A}\ }\textbf {\bibinfo {volume} {30}},\ \bibinfo {pages}
  {6525} (\bibinfo {year} {1997})}\BibitemShut {NoStop}%
\bibitem [{\citenamefont {Piekarewicz}\ \emph {et~al.}(2016)\citenamefont
  {Piekarewicz}, \citenamefont {Linero}, \citenamefont {Giuliani},\ and\
  \citenamefont {Chicken}}]{Piekarewicz:2016vbn}%
  \BibitemOpen
  \bibfield  {author} {\bibinfo {author} {\bibfnamefont {J.}~\bibnamefont
  {Piekarewicz}}, \bibinfo {author} {\bibfnamefont {A.~R.}\ \bibnamefont
  {Linero}}, \bibinfo {author} {\bibfnamefont {P.}~\bibnamefont {Giuliani}},\
  and\ \bibinfo {author} {\bibfnamefont {E.}~\bibnamefont {Chicken}},\
  }\bibfield  {title} {\bibinfo {title} {{Power of two: Assessing the impact of
  a second measurement of the weak-charge form factor of $^{208}$Pb}},\ }\href
  {https://doi.org/10.1103/PhysRevC.94.034316} {\bibfield  {journal} {\bibinfo
  {journal} {Phys. Rev. C}\ }\textbf {\bibinfo {volume} {94}},\ \bibinfo
  {pages} {034316} (\bibinfo {year} {2016})},\ \Eprint
  {https://arxiv.org/abs/1604.07799} {arXiv:1604.07799 [nucl-th]} \BibitemShut
  {NoStop}%
\bibitem [{\citenamefont {Beiner}\ \emph {et~al.}(1975)\citenamefont {Beiner},
  \citenamefont {Flocard}, \citenamefont {van Giai},\ and\ \citenamefont
  {Quentin}}]{Beiner:1974gc}%
  \BibitemOpen
  \bibfield  {author} {\bibinfo {author} {\bibfnamefont {M.}~\bibnamefont
  {Beiner}}, \bibinfo {author} {\bibfnamefont {H.}~\bibnamefont {Flocard}},
  \bibinfo {author} {\bibfnamefont {N.}~\bibnamefont {van Giai}},\ and\
  \bibinfo {author} {\bibfnamefont {P.}~\bibnamefont {Quentin}},\ }\bibfield
  {title} {\bibinfo {title} {{Nuclear ground state properties and
  self-consistent calculations with the Skyrme interactions: 1. Spherical
  description}},\ }\href {https://doi.org/10.1016/0375-9474(75)90338-3}
  {\bibfield  {journal} {\bibinfo  {journal} {Nucl. Phys. A}\ }\textbf
  {\bibinfo {volume} {238}},\ \bibinfo {pages} {29} (\bibinfo {year}
  {1975})}\BibitemShut {NoStop}%
\bibitem [{\citenamefont {Todd-Rutel}\ and\ \citenamefont
  {Piekarewicz}(2005)}]{ToddRutel:2005zz}%
  \BibitemOpen
  \bibfield  {author} {\bibinfo {author} {\bibfnamefont {B.~G.}\ \bibnamefont
  {Todd-Rutel}}\ and\ \bibinfo {author} {\bibfnamefont {J.}~\bibnamefont
  {Piekarewicz}},\ }\bibfield  {title} {\bibinfo {title} {{Neutron-Rich Nuclei
  and Neutron Stars: A New Accurately Calibrated Interaction for the Study of
  Neutron-Rich Matter}},\ }\href
  {https://doi.org/10.1103/PhysRevLett.95.122501} {\bibfield  {journal}
  {\bibinfo  {journal} {Phys. Rev. Lett.}\ }\textbf {\bibinfo {volume} {95}},\
  \bibinfo {pages} {122501} (\bibinfo {year} {2005})},\ \Eprint
  {https://arxiv.org/abs/nucl-th/0504034} {arXiv:nucl-th/0504034} \BibitemShut
  {NoStop}%
\bibitem [{\citenamefont {Fattoyev}\ \emph {et~al.}(2010)\citenamefont
  {Fattoyev}, \citenamefont {Horowitz}, \citenamefont {Piekarewicz},\ and\
  \citenamefont {Shen}}]{Fattoyev:2010mx}%
  \BibitemOpen
  \bibfield  {author} {\bibinfo {author} {\bibfnamefont {F.}~\bibnamefont
  {Fattoyev}}, \bibinfo {author} {\bibfnamefont {C.}~\bibnamefont {Horowitz}},
  \bibinfo {author} {\bibfnamefont {J.}~\bibnamefont {Piekarewicz}},\ and\
  \bibinfo {author} {\bibfnamefont {G.}~\bibnamefont {Shen}},\ }\bibfield
  {title} {\bibinfo {title} {{Relativistic effective interaction for nuclei,
  giant resonances, and neutron stars}},\ }\href
  {https://doi.org/10.1103/PhysRevC.82.055803} {\bibfield  {journal} {\bibinfo
  {journal} {Phys. Rev. C}\ }\textbf {\bibinfo {volume} {82}},\ \bibinfo
  {pages} {055803} (\bibinfo {year} {2010})},\ \Eprint
  {https://arxiv.org/abs/1008.3030} {arXiv:1008.3030 [nucl-th]} \BibitemShut
  {NoStop}%
\bibitem [{\citenamefont {Lalazissis}\ \emph {et~al.}(1997)\citenamefont
  {Lalazissis}, \citenamefont {Konig},\ and\ \citenamefont
  {Ring}}]{Lalazissis:1996rd}%
  \BibitemOpen
  \bibfield  {author} {\bibinfo {author} {\bibfnamefont {G.~A.}\ \bibnamefont
  {Lalazissis}}, \bibinfo {author} {\bibfnamefont {J.}~\bibnamefont {Konig}},\
  and\ \bibinfo {author} {\bibfnamefont {P.}~\bibnamefont {Ring}},\ }\bibfield
  {title} {\bibinfo {title} {{A new parametrization for the Lagrangian density
  of relativistic mean field theory}},\ }\href
  {https://doi.org/10.1103/PhysRevC.55.540} {\bibfield  {journal} {\bibinfo
  {journal} {Phys. Rev. C}\ }\textbf {\bibinfo {volume} {55}},\ \bibinfo
  {pages} {540} (\bibinfo {year} {1997})},\ \Eprint
  {https://arxiv.org/abs/nucl-th/9607039} {arXiv:nucl-th/9607039} \BibitemShut
  {NoStop}%
\bibitem [{\citenamefont {Fattoyev}\ and\ \citenamefont
  {Piekarewicz}(2013)}]{Fattoyev:2013yaa}%
  \BibitemOpen
  \bibfield  {author} {\bibinfo {author} {\bibfnamefont {F.~J.}\ \bibnamefont
  {Fattoyev}}\ and\ \bibinfo {author} {\bibfnamefont {J.}~\bibnamefont
  {Piekarewicz}},\ }\bibfield  {title} {\bibinfo {title} {{Has a thick neutron
  skin in ${}^{208}$Pb been ruled out?}},\ }\href
  {https://doi.org/10.1103/PhysRevLett.111.162501} {\bibfield  {journal}
  {\bibinfo  {journal} {Phys. Rev. Lett.}\ }\textbf {\bibinfo {volume} {111}},\
  \bibinfo {pages} {162501} (\bibinfo {year} {2013})},\ \Eprint
  {https://arxiv.org/abs/1306.6034} {arXiv:1306.6034 [nucl-th]} \BibitemShut
  {NoStop}%
\bibitem [{\citenamefont {Chabanat}\ \emph {et~al.}(1998)\citenamefont
  {Chabanat}, \citenamefont {Bonche}, \citenamefont {Haensel}, \citenamefont
  {Meyer},\ and\ \citenamefont {Schaeffer}}]{Chabanat:1997un}%
  \BibitemOpen
  \bibfield  {author} {\bibinfo {author} {\bibfnamefont {E.}~\bibnamefont
  {Chabanat}}, \bibinfo {author} {\bibfnamefont {P.}~\bibnamefont {Bonche}},
  \bibinfo {author} {\bibfnamefont {P.}~\bibnamefont {Haensel}}, \bibinfo
  {author} {\bibfnamefont {J.}~\bibnamefont {Meyer}},\ and\ \bibinfo {author}
  {\bibfnamefont {R.}~\bibnamefont {Schaeffer}},\ }\bibfield  {title} {\bibinfo
  {title} {{A Skyrme parametrization from subnuclear to neutron star densities.
  2. Nuclei far from stablities}},\ }\href
  {https://doi.org/10.1016/S0375-9474(98)00180-8} {\bibfield  {journal}
  {\bibinfo  {journal} {Nucl. Phys. A}\ }\textbf {\bibinfo {volume} {635}},\
  \bibinfo {pages} {231} (\bibinfo {year} {1998})},\ \bibinfo {note} {[Erratum:
  Nucl.Phys.A 643, 441--441 (1998)]}\BibitemShut {NoStop}%
\bibitem [{\citenamefont {{De Vries}}\ \emph {et~al.}(1987)\citenamefont {{De
  Vries}}, \citenamefont {{De Jager}},\ and\ \citenamefont {{De
  Vries}}}]{DEVRIES1987495}%
  \BibitemOpen
  \bibfield  {author} {\bibinfo {author} {\bibfnamefont {H.}~\bibnamefont {{De
  Vries}}}, \bibinfo {author} {\bibfnamefont {C.}~\bibnamefont {{De Jager}}},\
  and\ \bibinfo {author} {\bibfnamefont {C.}~\bibnamefont {{De Vries}}},\
  }\bibfield  {title} {\bibinfo {title} {Nuclear charge-density-distribution
  parameters from elastic electron scattering},\ }\href
  {https://doi.org/https://doi.org/10.1016/0092-640X(87)90013-1} {\bibfield
  {journal} {\bibinfo  {journal} {At. Data Nucl. Data Tables}\ }\textbf
  {\bibinfo {volume} {36}},\ \bibinfo {pages} {495} (\bibinfo {year}
  {1987})}\BibitemShut {NoStop}%
\bibitem [{\citenamefont {Erler}\ and\ \citenamefont
  {Su}(2013)}]{Erler:2013xha}%
  \BibitemOpen
  \bibfield  {author} {\bibinfo {author} {\bibfnamefont {J.}~\bibnamefont
  {Erler}}\ and\ \bibinfo {author} {\bibfnamefont {S.}~\bibnamefont {Su}},\
  }\bibfield  {title} {\bibinfo {title} {{The weak neutral current}},\ }\href
  {https://doi.org/10.1016/j.ppnp.2013.03.004} {\bibfield  {journal} {\bibinfo
  {journal} {Prog. Part. Nucl. Phys.}\ }\textbf {\bibinfo {volume} {71}},\
  \bibinfo {pages} {119} (\bibinfo {year} {2013})},\ \Eprint
  {https://arxiv.org/abs/1303.5522} {arXiv:1303.5522 [hep-ph]} \BibitemShut
  {NoStop}%
\bibitem [{\citenamefont {Gorchtein}\ and\ \citenamefont
  {Horowitz}(2009)}]{Gorchtein:2008px}%
  \BibitemOpen
  \bibfield  {author} {\bibinfo {author} {\bibfnamefont {M.}~\bibnamefont
  {Gorchtein}}\ and\ \bibinfo {author} {\bibfnamefont {C.~J.}\ \bibnamefont
  {Horowitz}},\ }\bibfield  {title} {\bibinfo {title} {{Dispersion gamma Z-box
  correction to the weak charge of the proton}},\ }\href
  {https://doi.org/10.1103/PhysRevLett.102.091806} {\bibfield  {journal}
  {\bibinfo  {journal} {Phys. Rev. Lett.}\ }\textbf {\bibinfo {volume} {102}},\
  \bibinfo {pages} {091806} (\bibinfo {year} {2009})},\ \Eprint
  {https://arxiv.org/abs/0811.0614} {arXiv:0811.0614 [hep-ph]} \BibitemShut
  {NoStop}%
\bibitem [{\citenamefont {Gorchtein}\ \emph {et~al.}(2011)\citenamefont
  {Gorchtein}, \citenamefont {Horowitz},\ and\ \citenamefont
  {Ramsey-Musolf}}]{Gorchtein:2011mz}%
  \BibitemOpen
  \bibfield  {author} {\bibinfo {author} {\bibfnamefont {M.}~\bibnamefont
  {Gorchtein}}, \bibinfo {author} {\bibfnamefont {C.~J.}\ \bibnamefont
  {Horowitz}},\ and\ \bibinfo {author} {\bibfnamefont {M.~J.}\ \bibnamefont
  {Ramsey-Musolf}},\ }\bibfield  {title} {\bibinfo {title} {{Model-dependence
  of the $\gamma Z$ dispersion correction to the parity-violating asymmetry in
  elastic $ep$ scattering}},\ }\href
  {https://doi.org/10.1103/PhysRevC.84.015502} {\bibfield  {journal} {\bibinfo
  {journal} {Phys. Rev. C}\ }\textbf {\bibinfo {volume} {84}},\ \bibinfo
  {pages} {015502} (\bibinfo {year} {2011})},\ \Eprint
  {https://arxiv.org/abs/1102.3910} {arXiv:1102.3910 [nucl-th]} \BibitemShut
  {NoStop}%
\bibitem [{\citenamefont {Erler}\ and\ \citenamefont
  {Gorchtein}(2020)}]{nuclearGZ}%
  \BibitemOpen
  \bibfield  {author} {\bibinfo {author} {\bibfnamefont {J.}~\bibnamefont
  {Erler}}\ and\ \bibinfo {author} {\bibfnamefont {M.}~\bibnamefont
  {Gorchtein}},\ }\href@noop {} {}\bibinfo {howpublished} {(private
  communication)} (\bibinfo {year} {Oct 2020})\BibitemShut {NoStop}%
\bibitem [{\citenamefont {Reed}\ \emph {et~al.}(2020)\citenamefont {Reed},
  \citenamefont {Jaffe}, \citenamefont {Horowitz},\ and\ \citenamefont
  {Sfienti}}]{Reed:2020fdf}%
  \BibitemOpen
  \bibfield  {author} {\bibinfo {author} {\bibfnamefont {B.}~\bibnamefont
  {Reed}}, \bibinfo {author} {\bibfnamefont {Z.}~\bibnamefont {Jaffe}},
  \bibinfo {author} {\bibfnamefont {C.~J.}\ \bibnamefont {Horowitz}},\ and\
  \bibinfo {author} {\bibfnamefont {C.}~\bibnamefont {Sfienti}},\ }\bibfield
  {title} {\bibinfo {title} {{Measuring the surface thickness of the weak
  charge density of nuclei}},\ }\href
  {https://doi.org/10.1103/PhysRevC.102.064308} {\bibfield  {journal} {\bibinfo
   {journal} {Phys. Rev. C}\ }\textbf {\bibinfo {volume} {102}},\ \bibinfo
  {pages} {064308} (\bibinfo {year} {2020})},\ \Eprint
  {https://arxiv.org/abs/2009.06664} {arXiv:2009.06664 [nucl-th]} \BibitemShut
  {NoStop}%
\bibitem [{\citenamefont {Horowitz}\ \emph {et~al.}(2012)\citenamefont
  {Horowitz} \emph {et~al.}}]{Horowitz:2012tj}%
  \BibitemOpen
  \bibfield  {author} {\bibinfo {author} {\bibfnamefont {C.~J.}\ \bibnamefont
  {Horowitz}} \emph {et~al.},\ }\bibfield  {title} {\bibinfo {title} {{Weak
  charge form factor and radius of $^{208}$Pb through parity violation in
  electron scattering}},\ }\href {https://doi.org/10.1103/PhysRevC.85.032501}
  {\bibfield  {journal} {\bibinfo  {journal} {Phys. Rev. C}\ }\textbf {\bibinfo
  {volume} {85}},\ \bibinfo {pages} {032501(R)} (\bibinfo {year} {2012})},\
  \Eprint {https://arxiv.org/abs/1202.1468} {arXiv:1202.1468 [nucl-ex]}
  \BibitemShut {NoStop}%
\bibitem [{\citenamefont {Reed}\ \emph {et~al.}(2021)\citenamefont {Reed},
  \citenamefont {Fattoyev}, \citenamefont {Horowitz},\ and\ \citenamefont
  {Piekarewicz}}]{Reed:2021nqk}%
  \BibitemOpen
  \bibfield  {author} {\bibinfo {author} {\bibfnamefont {B.~T.}\ \bibnamefont
  {Reed}}, \bibinfo {author} {\bibfnamefont {F.}~\bibnamefont {Fattoyev}},
  \bibinfo {author} {\bibfnamefont {C.~J.}\ \bibnamefont {Horowitz}},\ and\
  \bibinfo {author} {\bibfnamefont {J.}~\bibnamefont {Piekarewicz}},\
  }\href@noop {} {\bibinfo {title} {{Implications of PREX-II on the equation of
  state of neutron-rich matter}}} (\bibinfo {year} {2021}),\ \Eprint
  {https://arxiv.org/abs/2101.03193} {arXiv:2101.03193 [nucl-th]} \BibitemShut
  {NoStop}%
\bibitem [{\citenamefont {Riordan}\ \emph {et~al.}(2013)\citenamefont {Riordan}
  \emph {et~al.}}]{crex-proposal}%
  \BibitemOpen
  \bibfield  {author} {\bibinfo {author} {\bibfnamefont {S.}~\bibnamefont
  {Riordan}} \emph {et~al.} (\bibinfo {collaboration} {CREX Collaboration}),\
  }\href {https://hallaweb.jlab.org/parity/prex/c-rex2013_v7.pdf} {\emph
  {\bibinfo {title} {{Parity Violating Measurement of the Weak Charge
  Distribution of 48Ca to 0.02 fm Accuracy}}}},\ \bibinfo {type} {Tech. Rep.}\
  \bibinfo {number} {JLAB-PR-40-12-004}\ (\bibinfo  {institution} {TJNAF},\
  \bibinfo {year} {2013})\BibitemShut {NoStop}%
\bibitem [{\citenamefont {Benesch}\ \emph {et~al.}(2014)\citenamefont {Benesch}
  \emph {et~al.}}]{Benesch:2014bas}%
  \BibitemOpen
  \bibfield  {author} {\bibinfo {author} {\bibfnamefont {J.}~\bibnamefont
  {Benesch}} \emph {et~al.} (\bibinfo {collaboration} {MOLLER Collaboration}),\
  }\href@noop {} {\bibinfo {title} {{The MOLLER Experiment: An Ultra-Precise
  Measurement of the Weak Mixing Angle Using M\o ller Scattering}}} (\bibinfo
  {year} {2014}),\ \Eprint {https://arxiv.org/abs/1411.4088} {arXiv:1411.4088
  [nucl-ex]} \BibitemShut {NoStop}%
\bibitem [{\citenamefont {Souder}\ \emph {et~al.}(2008)\citenamefont {Souder}
  \emph {et~al.}}]{solid-PVDIS}%
  \BibitemOpen
  \bibfield  {author} {\bibinfo {author} {\bibfnamefont {P.}~\bibnamefont
  {Souder}} \emph {et~al.} (\bibinfo {collaboration} {SoLID}),\ }\href
  {https://hallaweb.jlab.org/collab/PAC/PAC34/PR-09-012-pvdis.pdf} {\emph
  {\bibinfo {title} {{Precision Measurement of Parity-violation in Deep
  Inelastic Scattering Over a Broad Kinematic Range}}}},\ \bibinfo {type}
  {Tech. Rep.}\ \bibinfo {number} {JLAB-PR-09-012-pvdis}\ (\bibinfo
  {institution} {TJNAF},\ \bibinfo {year} {2008})\BibitemShut {NoStop}%
\bibitem [{\citenamefont {Becker}\ \emph {et~al.}(2018)\citenamefont {Becker}
  \emph {et~al.}}]{Becker:2018ggl}%
  \BibitemOpen
  \bibfield  {author} {\bibinfo {author} {\bibfnamefont {D.}~\bibnamefont
  {Becker}} \emph {et~al.},\ }\bibfield  {title} {\bibinfo {title} {{The P2
  experiment}},\ }\href {https://doi.org/10.1140/epja/i2018-12611-6} {\bibfield
   {journal} {\bibinfo  {journal} {The Eur. Phys. J. A}\ }\textbf {\bibinfo
  {volume} {54}},\ \bibinfo {pages} {208} (\bibinfo {year} {2018})},\ \Eprint
  {https://arxiv.org/abs/1802.04759} {arXiv:1802.04759 [nucl-ex]} \BibitemShut
  {NoStop}%
\end{thebibliography}%

\end{document}